\documentclass[preprint,natbib]{aastex}
\usepackage{csvsimple}
\usepackage{adjustbox}
\usepackage{amsmath}
\shorttitle{}
\shortauthors{Herbst \& Greenwood}

\begin{document}

\title{A Radiative Heating Model for Chondrule and Chondrite Formation}

\author{William Herbst }
\affil{Astronomy Department, Wesleyan University,
    Middletown, CT 06459}
\author{James P. Greenwood}
\affil{Earth \& Environmental Sciences Department, Wesleyan University,
    Middletown, CT 06459}

\begin{abstract}

We propose that chondrules and chondrites formed together during a brief radiative heating event caused by the close encounter of a small (m to km-scale), primitive planetesimal (SPP) with incandescent lava on the surface of a large (100 km-scale) differentiated planetesimal (LDP). In our scenario, chondrite lithification occurs by hot isostatic pressing (HIP) simultaneously with chondrule formation, in accordance with the constraints of complementarity and cluster chondrites. Thermal models of LDPs formed near t=0 predict that there will be a very narrow window of time, coincident with the chondrule formation epoch, during which crusts are thin enough to frequently rupture by impact, volcanism and/or crustal foundering, releasing hot magma to their surfaces. The heating curves we calculate are more gradual and symmetric than the ``flash heating" characteristic of nebular models, but in agreement with the constraints of experimental petrology. The SPP itself is a plausible source of the excess O, Na and Si vapor pressure (compared to a solar nebula environment) that is required by chondrule observations. Laboratory experiments demonstrate that FeO-poor porphyritic olivine chondrules, the most voluminous type of chondrule, can be made using heating and cooling curves predicted by the ``flyby" model. If chondrules are a by-product of chondrite lithification, then their high volume abundance within well-lithified chondritic material is not evidence that they were once widespread within the Solar System. Relatively rare events, such as the flybys modeled here, could account for their abundance in the meteorite record.
 
\end{abstract}




\section{Introduction}

Chondrules are mm-scale igneous spherules, first described by \citet{Sorby1877}, which are a major component of most classes of the primitive meteorites named after them -- the chondrites \citep{Grossman1988}. Their origin remains as much a mystery in the eyes of most cosmochemists today as it was when they were discovered \citep{Ciesla2005, Palme2014,Connolly2016}. Laboratory experiments have shown that chondrule textures can be reproduced by melting pre-chondrule material at temperatures of 1400--1600~$^\circ$C for times of order 20 minutes and then cooling it on a time scale of hours or less \citep{Hewins1990,Desch2012}. Their spherical nature suggests this was done in space, or on an object with very small surface gravity. Their ages, measured by radioactive dating techniques, show that most chondrules were formed at t = 1--4 Myr, where t = 0 is set by the age of the most primitive solids (Ca/Al-rich inclusions; CAIs), also found in chondrites \citep{Villeneuve2009, Kita2013, Budde2016, JamesN.ConnellyMartinBizzarroAlexanderN.KrotAkeNordlundDanielWielandt2012, Connelly2017}.    

For decades, the most popular idea of chondrule formation has been that pre-existing dust aggregates embedded in the ``solar nebula", a gaseous remnant of the accretion disk through which the Sun formed, were heated and cooled by interaction with the gas (primarily H and He) that comprised the disk. This is known as the ``nebular hypothesis" \citep{Ciesla2005} and exists in a variety of forms today that differ according to the mechanism postulated to heat the gas. An {\it ad hoc} heating mechanism must be invoked because models of protoplanetary disk evolution, supported by observations, predict temperatures that are far below those required to form chondrules \citep{Chiang2010}. Proposed mechanisms include large-scale shocks from protosolar outburts or accreting clouds, more localized shocks from giant protoplanets on elliptical orbits and heating {\it via} lightning discharges \citep{Desch2012}. 

Within the last decade the nebular model has been challenged, and lost some of its popularity, as it has become clearer that the gaseous conditions under which chondrules formed were not at all representative of what one would expect in an accretion disk. In particular, the gas is inferred to be five orders of magnitude more oxidizing and has a much higher vapor pressure of Na than is plausible for nebular gas of solar composition \citep{Grossman2008, Alexander2008, Fedkin2012, Fedkin2016}. These discoveries have revived an older idea that chondrules might have formed as drops of ``fiery rain", to quote \citet{Sorby1877}, from splashes of hot liquid created \citep{Johnson} or released \citep{Sanders2012} during collisions of planetesimals. This class of models is known as ``planetary" and its primary attraction is that it may account for the gaseous environment in which chondrules are inferred to have formed. A collision model has also been invoked to explain unusual features of the CB chondrites \citep{Krot}. A major issue with planetary models, in general, is that no one has demonstrated with detailed modeling or experiments that the distinctive, often porphyritic, textures of most chondrules could be generated in such a manner. Their variable $\Delta^{17}$O is also inconsistent with formation from planetary interior melts \citep{Marrocchi2018}.

Additional arguments can be raised against both nebular and planetary models when the narrow time frame inferred for chondrule formation, t = 1--4 Myr, is considered. Accretion disks are expected and observed to dissipate monotonically with time and solar nebula gas may have disappeared entirely from the asteroid belt before the epoch of chondrule formation even began. \citet{Mamajek2009} finds a half-life for protoplanetary disks in local star forming regions of $\sim$1.7 Myr when all stellar masses are considered, but also evidence that disks last longer in very low mass stars, which shortens the estimate for stars near 1 M$_{\odot}$ to $\lesssim$1 Myr. \citet{Haisch2001} and \citet{Venuti2016} report that in the young cluster NGC 2264 more than half of the stars in their sample, most of which are of lower mass than the Sun, have completely lost their accretion disks within $\sim$3 Myr. In one star system, known as KH 15D, with an age of 3 Myr and a total mass of 1.4 solar masses, fortuitous geometry allows us to probe its protoplanetary disk at $\sim$3 AU; we find abundant solids in a vertically thin ring with no evidence for accompanying nebular gas \citep{Lawler2010, Aronow2017}. Furthermore, to explain the time gap between t = 0 and the primary epoch of chondrule formation, around t = 2 -- 3 Myr, one must invoke, in nebular models, a delayed gas heating mechanism. Among planetary models, only the proposal by \citet{Sanders2012}, which involves collisions to release (not create) liquid material, appears to address this issue.

A criticism of all chondrule formation theories to date is that, while they may account for the thermal history required by experimental petrology, they ignore other known constraints on chondrules and chondrites \citep{Connolly2016}. The work described here is motivated by the constraints of complementarity \citep{Wood1985, Hezel2008, Hezel2010, Palme2015, Ebel2016} and cluster chondrites \citep{Metzler2012}, that chondrule and chondrite formation were closely linked in space and time. We consider a model in which they were simultaneous. Complementarity is the label associated with the observation that while whole chondrites may have a chemical abundance pattern that closely follows the composition of the Sun, their two primary components, chondrules and matrix, do not. Underrepresentation of an element in one component is compensated for by overrepresentation in the other component, implying formation of the chondrite in a closed system. The phenomenon of complementarity has recently become an even more powerful constraint as a result of the discovery of nucleosynthetic complementarity in tungsten isotopes between chondrules and matrix in several carbonaceous chondrites \citep{Becker2015, Budde2016}.  Cluster chondrites are found in all groups of unequilibrated ordinary chondrites. They are regions where chondrules of a variety of textures are closely packed and deformed in a manner indicating that the chondrite formed while the chondrules were still hot. Again the inference is that there was close spatial and temporal coincidence of chondrule and chondrite formation. 

As is the case with chondrules, there is at present no consensus among cosmochemists on how, when or where chondrite lithification occurs. \citet{Consolmagno1999} describes some potential mechanisms and issues. One requires a source of elevated pressure and/or temperature to reduce the porosity of the material. Terrestrial rocks offer little guidance because they form under steady pressures of a magnitude that cannot be achieved on asteroids, even Ceres. Transient pressure events due to collisional impacts are often invoked as the primary mechanism for chondrite lithification, although models differ in their details \citep{Beitz2016, Lichtenberg2018}. Recently, the hot isostatic pressing (HIP) process, involving heat and mild pressure, has been invoked, but at lower temperatures and much longer timescales than are employed commercially \citep{Atkinson2000, Gail2015}. None of these studies address the challenges of complementarity, cluster chondrites or other links between chondrules and their host chondrites, such as the size-group relationship \citep{Friedrich2015}. Chondrule formation and chondrite lithification are currently modeled independently of one another.        

In this paper we propose a model for simultaneous chondrule and chondrite formation. It employs radiative heating and predicts temperature curves that are consistent with the constraints of experimental petrology. It is an extension of the ``flyby model", originally proposed by \citet{Herbst2016} (hereinafter Paper I) for chondrule formation, but now explicitly considers the heating of a larger object characterized by certain bulk parameters including opacity. In Section 2 we provide a basic overview of the model and its components. In Section 3 we develop a more comprehensive theory of the heating based on a gray, plane parallel solution to the equation of radiative transfer. We consider the full range of possible orbital parameters for the flyby from circular to hyperbolic. In Section 4 we discuss preliminary laboratory work aimed at testing the distinctive thermal histories that the flyby model predicts and demonstrate that objects with chondrule textures will form under relevant conditions. Section 5 is a discussion of some of the implications, challenges and potential tests of this model, including what happens to the chondrite after it forms. We follow \citet{Elkins-tanton2011} in assuming that chondritic material gradually accretes to LDPs creating an undifferentiated crust where metamorphisis can occur. Our contribution to the scenario is that chondrites, with their complement of chondrules, arrive on the LDPs in already lithified form as part of that chondritic material. Section 6 is a brief summary of the paper.

\section{Overview of the Flyby Model}

A schematic representation of our model is shown in Fig. 1. The small primitive planetesimal (SPP) has been assembled and stored over 1-4 Myr from solids in its formation zone and is small enough (1-1000 m) to have avoided melting by $^{26}$Al. It may be a ``free floating" object within the protosolar disk, which encounters the larger planetesimal by chance, or a gravitationally bound, orbiting moonlet or ring particle. The large differentiated planetesimal (LDP) is sufficiently large ($\ge$100 km) and formed sufficiently early to have a molten interior due to $^{26}$Al. We assume that crustal foundering or rupture, possibly by impacts, will lead to the episodic appearance of incandescent lava at the object's surface. Infrared radiation from these lava eruptions at presumed temperatures near 2000 K serves as the heat source for both chondrule formation and chondrite lithification. We now consider these basic elements of the model in somewhat more detail.  

\subsection{The Heat Source: Exposed Magma on Large Differentiated Planetesimals (LDPs)}

Experimental petrology and radioactive dating of chondrules have provided strong constraints on the nature, timing and location of possible chondrule heat sources \citep{Hewins1990, Desch2012, Kita2012, Kita2013}. A plausible heat source for chondrules must be capable of raising the temperature of pre-chondrule material to about 1400 -- 1600~$^\circ$C for about 20 minutes or so, and not much, if any, beyond that \citep{Levy1988, Hewins1990}. It must also allow for cooling that is rapid, but moderated, over a time scale of minutes to hours. The heat source must have been active locally over a range of disk radii during the chondrule formation epoch of t = 1--4 Myr and must have shut down permanently thereafter to explain the absence of younger chondrules. Most cosmochemists agree that there is a significant gap in time between the formation of the CAI's and the formation of the first chondrules based on their $^{26}$Al ages \citep{Kita2005, Villeneuve2009, Budde2016}, although there is certainly debate on this point \citep{JamesN.ConnellyMartinBizzarroAlexanderN.KrotAkeNordlundDanielWielandt2012, Connelly2017}.

A heating mechanism that meets all of these constraints is radiative heating by exposure to incandescent lava at the surfaces of LDPs during a close flyby \citep{Herbst2016}. Chondrule precursors exposed to the infrared radiation from a lava ocean in this manner can reach sub-liquidus or possibly liquidus temperatures, but cannot get hotter than the temperature of the lava ocean, in agreement with experimental constraints on the temperatures experienced by chondrules. The timescale for the most intense heating is naturally of the order of minutes, because that is the dynamical timescale of the gravitational interaction (see below). This mechanism could only work during a brief interval near the beginning of the solar system because it relies on the decay of $^{26}$Al to melt the LDPs over a large portion of their volume and the half-life of $^{26}$Al is only 717,000 years. The time gap between CAI formation and chondrule formation is the time required to form and fully melt the LDPs. We turn to a more quantitative assessment of these aspects of the flyby model. 

The existence of LDPs in the Solar System by t = 1 Myr is supported by the age of the iron meteorites, which are indistinguishable from the ages of the CAI's \citep{Scherste2006}. Evidently, bodies large enough to melt substantially and differentiate formed shortly after t = 0 within the solar nebula. The mechanism for such early planetesimal formation is still debated by astrophysicists \citep{Chiang2010}. It may have happened by a gravitational instability within a sub-layer of solids, which naturally produces objects of the requisite scale \citep{Goldreich1973}, or it may have been driven or assisted by a streaming instability, or other mechanisms related to the gas \citep{Chiang2010}. But the meteorite record leaves no doubt that large planetesimals formed rapidly, melted and differentiated very close to t~=~0. The inferred abundance of $^{26}$Al at t~=~0, is more than sufficient to be the cause of the melting, as first pointed out by \citet{Lee1977} and developed in substantial detail by \citet{Hevey2006}. The amount of energy required to fully melt one gram of primitive dust is about 1.6 kJ while the canonical abundance of $^{26}$Al provides about 6.4 kJ per gram at t~=~0 -- more than four times the amount needed to fully melt any assemblage large enough to be insulated. Any object with a radius exceeding about 50 km that forms within the first few hundred thousand years, would have become substantially liquid within only a few hundred thousand years additional ``incubation" period \citep{Hevey2006}.

In this picture, a gap between the time of CAI formation and chondrule formation is predicted, since it requires some time to form the large planetesimals and raise their temperatures to the liquidus throughout the interior. To illustrate the situation in more detail, we show, in Fig.~2, the evolution of the crustal thickness of planetesimals of three different sizes (r = 50, 100 and 200 km) formed at two different times (t$_f$ = 0 and t$_f$  = 0.75 Myr) and compare with the observed distribution of chondrule ages reported by \citet{Villeneuve2009}. The figure is based on the finite difference model of \citet{Hevey2006}, which includes convection in the interior. According to their model, the crust of a 100 km radius planetesimal that formed at t=0 would shrink to a mere 250 meters by t = 0.5 Myr. \citet{Sanders2012} have noted that the crust, being more dense than the magma flowing beneath it, would be susceptible to rupture by impact or convective drag, allowing the incandescent magma to reach the surface. They propose that chondrules could form directly from that magma; we propose that infrared radiation from the surface lava could heat a nearby SPP, leading to both chondrule and chondrite formation.

As Figure 2 shows, by t = 4 -- 5 Myr, crusts are expected to have thickened sufficiently that, even for the largest planetesimals, they will be difficult to rupture. If the LDP magma is the heat source for chondrule formation then one would expect a termination of the epoch of chondrule formation at about that time. We consider the fact that our proposed heat source explains all three aspects of the timing of chondrule formation -- the main epoch around 2 Myr, the gap between CAI formation and that epoch, and the termination of chondrule formation near 4 Myr -- to be a significant point in favor of it. The \citet{Sanders2012} planetary model shares this advantage with us, but other nebular and planetary models do not.  

\begin{figure}
\label{fig0}
\plotone {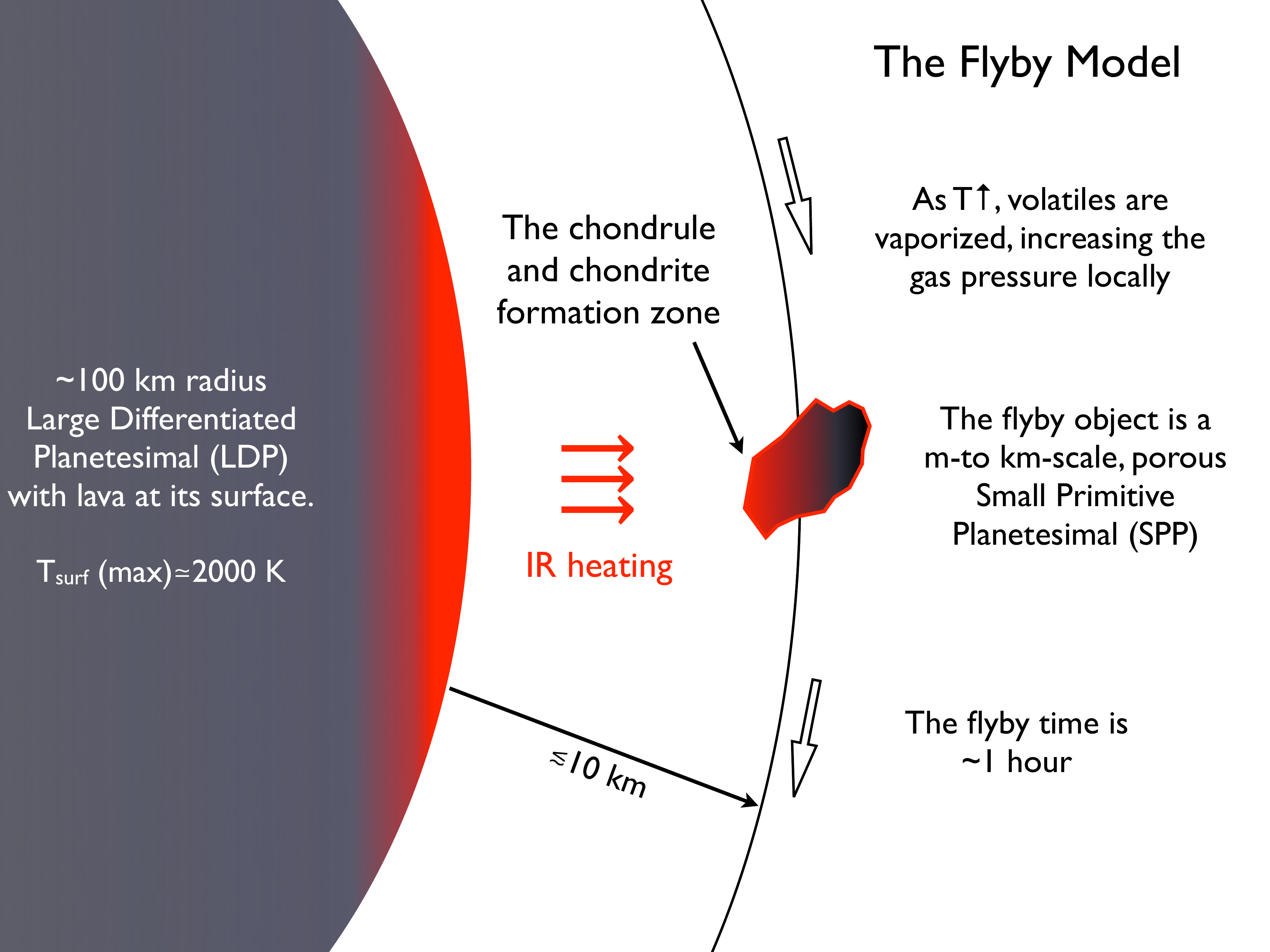}
\caption{A schematic representation of the flyby model. A large, early-formed, differentiated planetesimal (LDP, left) has molten lava at its surface as a small, primitive planetesimal (SPP) flys by within $\sim$10\% of the radius of the large object. Events such as these should be relatively common during the chondrule forming era of 1--4 Myr, but rare or absent at other times because the crusts of the LDPs thicken with time. The thermal history of material at the irradiated surface of the primitive planetesimal is consistent with the requirements of chondrule formation based on laboratory simulations. It depends only on the temperature of the extruded lava, the distance of closest approach and the density of the differentiated planetesimal. We propose that, simultaneous with chondrule formation, the irradiated surface of the primitive planetesimal experiences lithification to chondrite densities through hot isostatic pressing (HIP), as described further in the text.}
\end{figure}

\begin{figure}
\label{fig0}
\plotone {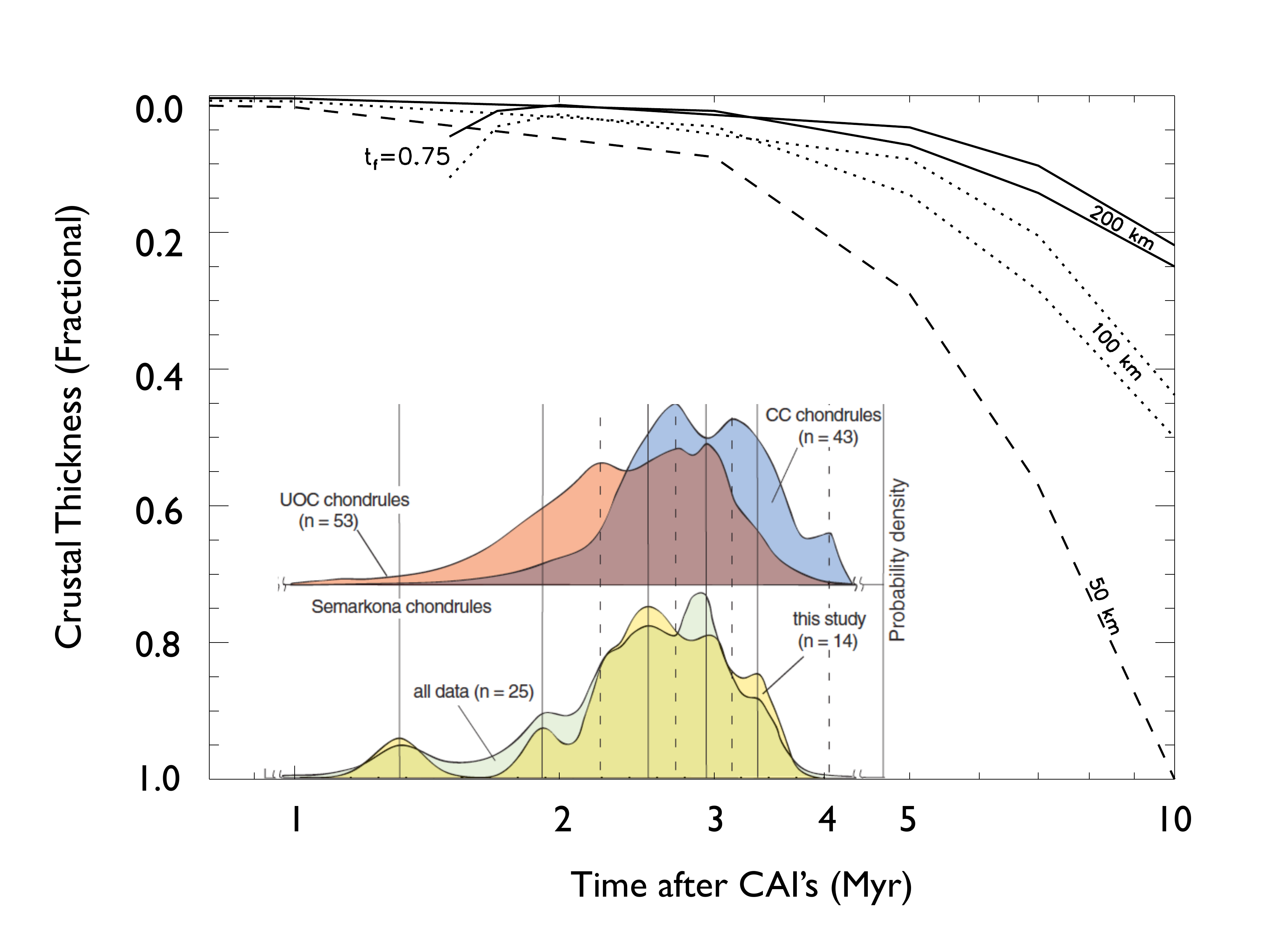}
\caption{The evolution of fractional crustal thickness of planetesimals of three different radii formed at two different times is shown, based on the thermal evolution models of \citet{Hevey2006}. This is compared to chondrule ages for carbonaceous chondrites (CC's) and ordinary chondrites (OC's) reported by \citet{Villeneuve2009}, as shown in that paper. It is clear that large planetesimals (r $>$ 100 km) forming early (t $<$ 0.75 Myr) in the history of the solar system will have extremely thin crusts at just the observed epoch of chondrule formation. The radiant energy, primarily at infrared wavelengths, released in episodic crustal ruptures is the heat source for both melting chondrules and lithifying chondrites in the flyby model.}
\end{figure}

\subsection{The Flyby Heating Timescale}

LDPs of initially chondritic composition will settle to a density of $\rho \approx 3.3$ gm cm$^{-3}$ \citep{Hevey2006}. To get hot enough, a chondrule precursor needs to pass within about 10\% of the radius of the planetesimal \citep{Herbst2016}. This implies a mean density of $ \bar \rho = 2.5 - 3.3 $ gm cm$^{-3}$ for the sphere interior to the flyby distance. A representative flyby timescale (t$_{fb}$) for gravitational interactions with objects of mean density $\bar \rho$ is one-half the circular orbital period, or
\begin{equation}
t_{fb} = {\sqrt{3 \pi \over {4 G \bar \rho}}}  
\end{equation} 
where G is the gravitational constant. For a typical value of $ \bar \rho = 3$ gm cm$^{-3}$, $t_{fb} = 57$ minutes. The characteristic time scale for dynamical interactions of the sort that could lead to appropriate heating in the flyby model is, therefore, naturally in the range required by experimental petrology. The variation of temperature with time experienced by any particular flyby object will vary according to other details of the event; this is explored in Section 3 of the paper. Important factors which we model in this paper include the angular extent of the exposed lava ocean, the eccentricity of the orbit in the gravitationally bound case, and the approach speed of the object if the orbit is open. Rotation of the planetesimal and of the flyby object may also affect heating times, but probably not by factors larger than a few; analysis of possible rotational effects is beyond the scope of this paper.  

\subsection{The Flyby Object: a Small Primitive Planetesimal (SPP)}

The formation and evolution of SPPs is a challenge to theorists \citep{Chiang2010} but they must have existed as steps on the way to larger objects. Recent 3-d simulations of the formation of primitive objects in a protostellar nebula indicate that SPPs will form easily and rapidly, on a timescale of thousands of years, and might grow to be as large as 100~m \citep{Garcia}. \citet{Kataoka2013} even chart a path for icy SPPs that leads to comet nuclei in the outer solar system, although that might be harder in the asteroid belt. In general, the products of these modeling efforts are extremely porous structures, with volume filling factors ($\phi$) of order 10$^{-4}$. After formation, the porosity will presumably be reduced to some extent by interaction with ambient gas or other aggregates. CAI's and pre-solar grains will be among the accreted constituents, alleviating any need to store them separately in the asteroid belt for 1 Myr or more.

We assume that many SPPs will not survive for the 1--4 Myr necessary to create the conditions for chondrule formation. They may accrete to LDPs prior to that time or suffer destructive collisions with each other, creating small grains that can be removed from the asteroid belt on short timescales by blow out or Poynting-Robertson drag. The ones of interest here are those that accrete onto LDPs late enough to avoid being subsumed by surface lava extrusions, thereby forming chondritic crusts, as described by \citet{Elkins-tanton2011}. A subset of those that accrete between 1--4 Myr will experience flyby heating in the manner proposed here either prior to or during accretion to the LDP. Some of them may also suffer metamorphism and shock events at later times, after accretion, but these events would be unrelated to the flyby heating responsible for their lithification and chondrule formation. We note that our model differs from the canonical view in one respect, namely, we propose that chondrites form in space, rather than on a ``parent body". But if lithification is unrelated to metamorphic or impact processing, this is not a significant difference with canoncial views, although it would suggest substitution of the term ``host body" for ``parent body".   

To experience sufficient heating an SPP must pass within about 10\% of the radius of an LDP with an active surface. This is well inside the Roche limit, so if the SPP is large enough to be held together by gravity before its flyby encounter, it will fragment into smaller objects during the first such close passage. An interesting possibility is that the SPPs may exist for a time as moonlets or ring particles of the LDPs, analogous to Saturn's rings. Rings of small SPPs may have accompanied the formation of the LDP or been generated later by the tidal disruption of a larger SPP during a close encounter. If the SPPs are gravitationally bound to the LDP at any point they can be stored indefinitely in the asteroid belt and their chance of experiencing flyby heating will be greatly enhanced. There is evidence that some chondrules have experienced multiple heating events \citep{Krot1997, Rubin2017}, which could also be explained if SPPs are in gravitationally bound orbits, but otherwise seems unlikely.   

\subsection{The Gaseous Environment of Chondrule and Chondrite Formation}

Cosmochemical studies have shown that the FeO content of chondrules requires formation in a gas that is five orders of magnitude more oxidizing than the solar nebula \citep{Grossman2008}. \citet{Fedkin2012} have shown that chondrules would exhibit unseen isotopic anomalies if they formed under standard nebular conditions. \citet{Alexander2008} concluded on the basis of Na abundance patterns in 26 chondrules that they must have formed in regions characterized by a density of solids in the range of 10$^{-3}$ to 10$^{-5}$ gm cm$^{-3}$, or possibly higher. If this region contained a full complement of H and He gas from the solar nebula it would be gravitationally unstable. The half-life for gaseous disks around 1 M$_{\odot}$ stars based on observations of young stellar clusters is $\lesssim$1 Myr \citep{Mamajek2009} and in the well-studied few Myr old cluster NGC 2264, less than half of the solar-like stars still show any evidence at all for accretion disks \citep{Venuti2016}. We have, therefore, not considered any effects that remaining H and He gas from the solar nebula might have on the dynamics of a flyby encounter.

Magmatic activity on the LDPs might result in the release of volatiles, creating a transient atmosphere. However, this gas will be quickly lost from the planetesimal, since the escape velocity from its surface is only 0.14 km s$^{-1}$ (taking the example of a 100 km radius planetesimal with a mean density of 3.3 gm cm$^{-3}$), while the root mean square speed of an Oxygen atom at T = 2000 K, for example, is 1.75 km s$^{-1}$. Within seconds of any eruption and gas release the volatiles would have left the region of interest. Hence, we believe the environment in which chondrule formation will proceed will be dominated by the environment that the flyby object creates for itself during the heating event. Ambient gas from the nebula or the large planetesimal responsible for its heating can probably be safely ignored in most instances, and we take that approach here. 

If we assume a grain density of $\rho_g = 1$ gm cm$^{-3}$, and a porosity of $\phi = 10^{-1} - 10^{-4}$, then the bulk density of the material, $\rho_m = \phi \rho_g$, will be in the range $10^{-1} - 10^{-4}$ gm cm$^{-3}$, which is consistent with the constraints reported by \citet{Alexander2008}. For the flyby model to be successful, the irradiated object must be a porous aggregate of chondrule precursors that is capable of retaining, at least partially (through a closed pore structure) its own gaseous environment during the heating episode. If our model is to account for the phenomenon of complementarity it will also be necessary to confine gas released during the melting of chondrules, so that it can recondense as matrix within the same meteorite, and this also requires a closed pore structure over at least some portion of the SPP. Finally, the HIP mechanism that we propose for lithification requires some degree of pressure, which we assume comes from volatiles comprising the SPP as they heat up and are at least partly trapped by a closed pore environment. 

\section{The Thermal Model}

Our thermal model for the heated surface of an SPP exposed during a close flyby to hot lava on the surface of an LDP (see Fig. 1) follows the gray plane layer solution to the equation of radiative transfer described in Chapter 11 of \citet{howell2016}. The assumptions of this solution are that the geometry is plane parallel, that the opacity ($\kappa$) is independent of wavelength (gray) or location within the layers and that the condition of radiative equilibrium applies, i.e. one neglects energy transport by convection or conduction. The boundaries form two surfaces of an enclosure of optical thickness $\tau_D$. In our application, significant radiation, characterized by a black body temperature, T$_1$, is incident on only one side of the layer, the side facing the hot lava (see Fig. 3). The other side (i.e. towards the inside of the planetesimal; upwards on Fig. 3) extends to a large optical depth. Since there is a finite amount of time and energy available to heat material during a flyby, thermal equilibrium will only be achieved to a finite depth, D, or optical depth $\tau_D$. The radiation incident on the top of the layer, T$_2$, from the interior of the planetesimal beyond D is neglected. With these assumptions, the equilibrium temperature distribution within the irradiated surface layer of the SPP, T($\tau$), follows from Equation (11.59) of \citet{howell2016}:
\begin{equation}
T^4 (\tau) = {{1 \over 2 \sigma} \Big[{\pi \int_0^1 I^+(0,\mu)e^{-{\tau \over \mu}} d\mu} + {\int_0^\infty \sigma T^4(\tau ')} E_1(|\tau ' - \tau|) d\tau '}\Big]
\end{equation}
where $\sigma$ is the Stefan-Boltzmann constant, $I^+ (0, \mu)$ is the incident intensity at the surface ($\tau = 0$) of the SPP as a function of $\mu$ = cos ($\theta$), $\theta$ being the angle between the normal to the surface and the direction of the intensity, and E$_1$ is one of the exponential integral relations defined by \citet{Chandrasekhar1960} (see Appendix D of \citet{howell2016}).

\begin{figure}
\label{fig0}
\plotone {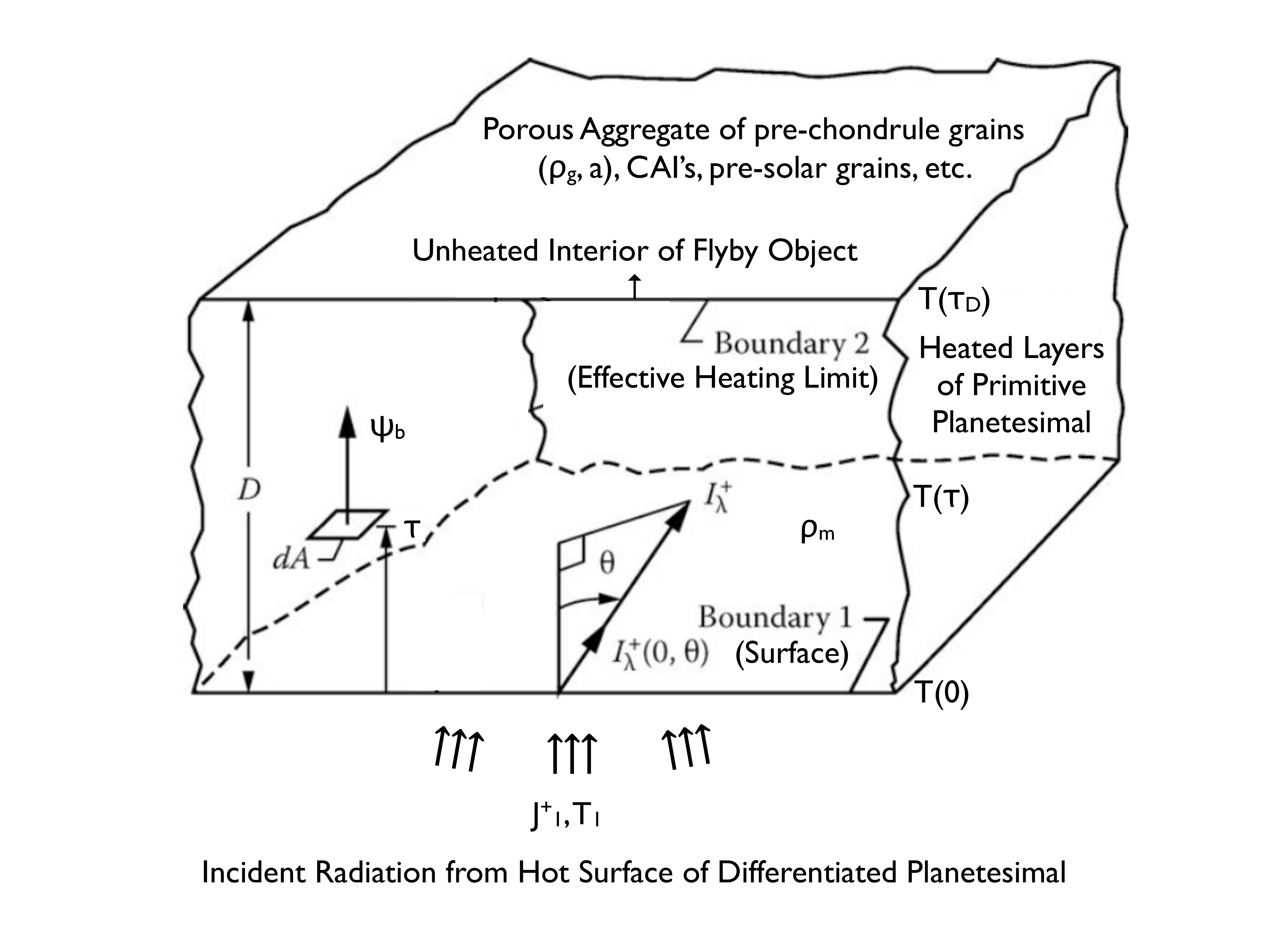}
\caption{The geometry of our model for the heating of the surface of an SPP during a close flyby of a hot source. The layer contains all of the ingredients of a chondrite and is characterized by a depth, D, that achieves thermal equilibrium, an opacity, $\kappa$, a mean density, $\rho_m$. and a Bond albedo, A. The opacity is primarily from grains of radius, a, and grain density, $\rho_g$. Infrared radiation is incident on the surface of the SPP from below and penetrates upward to an optical depth of $\tau_D$. An equilibrium temperature gradient is established between optical depths 0 and $\tau_D$ described by T($\tau$). A constant fraction $\Psi_b$ of the incident flux is transferred through the zone by radiation. This figure is adapted from Fig. 11.1 of \citet{howell2016}.}
\end{figure}

This solution shows explicitly that there are two components to the radiative flux responsible for heating the SPP. The first term on the right hand side of Equation 2 accounts for radiation incident on the surface from outside and the second term accounts for the radiation of the heated layers themselves. The temperature at any optical depth within the SPP, T($\tau$), may be expressed in terms of the dimensionless parameter $\phi_b$, defined as 
\begin{equation} 
\phi_b(\tau) = {{T^4(\tau) - T^4_2} \over {T^4_1 - T^4_2}}
\end{equation}
where the subscript b emphasizes that this applies in the case of black body boundaries. A convenient graphical representation of this result is shown in Fig. 11.4 of \citet{howell2016}, based on the calculations of \citet{Heaslet1965}.

T$_1$ is set by matching the actual flux of radiant energy incident on the SPP's surface to what a black body at T$_1$ would provide. We calculate the mean intensity incident in the positive direction, denoted J$_1^+$, as
\begin{equation}
J_1^+ \equiv {{1 \over {2 \pi}} \int I^+ d\Omega} = I_0 ({\Omega \over {2 \pi}}) (1 - A) 
\end{equation}
where $I_0$ is the surface intensity of the hot portion of the LDP, $\Omega$ is the solid angle subtended of that hot portion as viewed from the surface of the SPP and A is the Bond albedo. The equivalent black body temperature then follows as:
\begin{equation}
J_1^+ = {\sigma T_1^4 \over  \pi}.
 \end{equation}

The radiation from the hot portion of the LDP is also represented as a black body of surface temperature $T_s$, i.e. the temperature of the surface lava, hence 
\begin{equation}
I_0 = {\sigma T_s^4 \over \pi}.
\end{equation}
Combining expressions leads to
\begin{equation}
T_1^4 = T_s^4 ({\Omega \over {2 \pi}}) (1 - A). 
\end{equation}
With $T_2 = 0$, the expression for the temperature, T($\tau$) of the SPP as a function of optical depth into it, ($\tau$), reduces to
\begin{equation}
T(\tau) = T_s [{\phi_b(\tau) ({\Omega \over {2 \pi}}) (1 - A)}]^{1 \over 4}. 
\end{equation}
The temperature is highest, of course, at the surface of the SPP ($\tau = 0$) and the model thermal histories we calculate in Sections 3.1, 3.2 and 3.3 are for T(0), and depend only on $\phi_b(0)$. In section 3.4 we consider the full temperature profile within the irradiated surface. Table 1 summarizes the symbols, units and, where appropriate, values adopted in the calculations.

\begin{deluxetable}{ccccc}
\centering
\tablewidth{0pt}  
%\tablecaption{Derived Stellar Properties of KH 15D Components\tablenotemark{a} \label{table:derived_stellar_props}}
\tabletypesize{\scriptsize}
\tablecaption{Variables and Free Parameters of the Flyby Thermal Model}
\tablehead{\colhead{Quantity} & \colhead{Symbol} & \colhead{Units}& \colhead{Value} & \colhead{Reference}}
\startdata

Bond Albedo & A & dimensionless & 0.04 & \citep{Lamy2004}; see text \\
Lava Temperature & T$_s$ & Kelvins & 2000 & \cite{Hewins1990,DeKleer2014}\\
Blackbody Incident Temperature & T$_1$ & Kelvins & variable & Equation 7\\
Opacity & $\kappa$ & cm$^2$ gm$^{-1}$ & 10 & Equation 14\\
Optical thickness of layer & $\tau_D$ & dimensionless & 24 & Section 3.4 \\
Net Flux & $\Psi_b$ & dimensionless & 0.05 & \citet{Heaslet1965} \\
Temperature Jump & $\phi_b(0)$ & dimensionless & 0.93 & \citet{Heaslet1965} \\
Solid Angle Subtended by Lava Ocean & $\Omega$ & sterradians & variable & Sections 3.1, 3.2 \& 3.3\\
Optical Depth into Heated Layer & $\tau$ & dimensionless & variable & Equation 13\\
Temperture of Layer & T($\tau$) & Kelvins & variable & Equation 8\\
Grain Density & $\rho_g$ & gm cm$^{-3}$ & 3 & \citet{Friedrich2015} \\
Surface Density of SPP & $\rho_m$ & gm cm$^{-3}$ & 10$^{-1}$ - 10$^{-4}$ & \citet{Alexander2008} \\
Grain Radius & a & cm & 0.05 & \citet{Friedrich2015} \\
Absorption Efficiency & Q$_{abs}$ & dimensionless & variable & \citet{Draine2011}; Fig. 10 \\
Specific Heat of Grains & C & Joules kg$^{-1}$ K$^{-1}$ & 840 & \citet{Opeil2012} \\

\enddata
%\tablenotetext{a}{Note}
\end{deluxetable}

To calculate a thermal model for T(0) we must specify three parameters, the surface temperature of the hot lava on the LDP, $T_s$, the value of $\phi_b(0)$, and the Bond albedo, A, of the SPP. It is apparent that the results are more sensitive to $T_s$ than to $\phi_b(0)$ or A. The value of $\phi_b(0)$ varies depending on the full optical thickness of the medium enclosed between the boundaries ($\tau_D$) because there are two heat sources -- the incident radiation at the boundary facing the hot lava and the radiation from each layer within the medium. As the material heats up, initially by radiation from the external source alone, it quickly becomes, itself, an important component of the radiant energy field that sets the equilibrium temperature structure. For optically thin media, $\phi_b(0) \rightarrow 0.5$ as $\tau_D \rightarrow 0$. In this case, we recover the result employed in Paper I that $T(0) = 2^{1 \over 4} T_1$. For optically thick media $\phi_b(0) \rightarrow 1.0$ as $\tau_D \rightarrow \infty$ and, therefore, $T(0) \rightarrow T_1$. In our application, the surface layers of the SPPs are certainly optically thick, but the heating interval is relatively short (minutes to an hour) so one may question the depth within the SPP to which thermal equilibrium is established. We show below that one may expect a rapid thermal response from the material, on a timescale of seconds or less. For these reasons, we have adopted $\phi_b(0) = 0.93$, corresponding to $\tau_D = 10$, in calculating expected thermal histories during a flyby. Since $T(0) \propto \phi_b(0)^{1 \over 4}$ it is relatively insensitive to this choice. For $\tau_D = 2.0$, $ \phi_b(0) = 0.82$, so $T(0)$ would be reduced by only 3 percent. Most likely, $\tau_D$ will exceed 10; we estimate a value of 24 in Section 3.4, in which case $\phi_b(0)$ will be a bit larger than 0.93, but, again, there is not much effect on the values of T(0).

The Bond albedo, A, is the fraction of the incident radiant energy that is lost to the SPP by reflection from its surface. This is likely to be very small, given that the primitive planetesimal is extremely porous and the energy is transported primarily by near-infrared radiation. Perhaps the best solar system guides are comets, which are known to be generally dark \citep{Lamy2004}. Some recent, high precision, measurements of the Bond albedo of cometary nuclei include 67P/Churyumov-Gerasimenkov, $A = 0.019 \pm 0.001$ \citep{Statella2017}, and 103P/Hartley 2, $A = 0.012 \pm 0.002$ \citep{li2013}. For comparison, a recent high precision determination of the Bond albedo of Asteroid (21) Lutetia, $A = 0.076 \pm 0.002$, was reported by \citet{Masoumzadeh2015}. Based on these results we adopt a value of A = 0.04 for the thermal models presented here. As is the case with $\phi_b(0)$ it seems safe to assume that the results could not be altered by more than a percent or two by any alternate reasonable choice, given the low sensitivity of the surface temperature to this parameter.

The only other free parameter in our thermal model is the surface temperature, T$_s$, of the hot lava on the LDP. The liquidus temperature of silicates depends on their chemical composition and varies from about 1500 K to 2150 K for compositions characteristic of chondrules \citep{Herzberg1979,Hewins1990}. Of course the material could erupt onto the surface of the LDP at even higher temperatures. \citet{Hin2017} have proposed that large-scale vapor loss from growing planetesimals is the best way to account for their discovery that all differentiated bodies have isotopically heavier magnesium compositions than chondritic meteorites. They invoke vaporization by planetesimal collisions but it is also possible that overheating by $^{26}$Al decay could be involved. The only solar system guide that we have is Io, which is overheated by tidal interaction with Jupiter and erupts regularly. While most of its eruptions are inferred to be at lower temperatures, one (on 29 August 2013) was suggestive of temperatures 1900 K or higher \citep{DeKleer2014}, which are associated with ultramafic lava composition. In Paper I, we adopted $T_s = 2150$ K since the intent was to show that the flyby mechanism was possible. Here we adopt a slightly lower value of $T_s = 2000$ K. In spite of this, the thermal histories we calculate reach higher temperatures than displayed in Paper I because our thermal model now explicitly includes radiation from the heated layers of material below the surface, as described above.

\subsection{The Case of Fully Molten LDPs}

With T$_s$, $\phi_b(0)$ and A fixed it is clear from Equation 8 that the calculation of surface temperature, T(0), with time, t, reduces to the calculation of $\Omega(t)$. If an entire facing hemisphere of a spherical LDP of radius $R_p$ is molten, then we may write:
\begin{equation} 
\Omega(t) = \Omega_p(t) = 2 \pi ( 1 - cos\ \omega(t))
\end{equation} 
where $\omega(t) = {sin^{-1} ({R_p \over r(t)})}$ and $r(t)$ is the distance from the surface of the SPP to the center of mass of the LDP. $\Omega_p$ is the solid angle subtended by the LDP as seen from the surface of the SPP. This is a limiting case for our model, providing the maximum temperature that can be achieved. To calculate $\Omega (t)$ in this case, we first follow Paper I and adopt a parabolic orbit for the flyby. There are only two free parameters then, the mean density ($\rho$) of the LDP and the ratio of the distance of closest approach of the surface of the SPP to the radius, R$_p$, of the LDP. For definiteness we adopt $\rho = 3$ gm cm$^{-3}$ and $R_p$ = 100 km.

Fig. 4 shows the variation of surface temperature on the SPP with time for three different choices of closest approach: 2, 10 and 50 km. Again, one can scale the results to planetesimals (with $\rho = 3$ gm cm$^{-3}$) of any size simply by keeping the ratio of closest approach to the planetesimal radius constant. The time axis scales as the inverse square root of the mean density of the LDP. Hence, our thermal model has very limited flexibility to predict different heating curves. As noted in Paper I and evident in Fig. 3, a characteristic feature of thermal histories predicted by the flyby model is that they are smoothly varying and relatively flat-topped. This is in contrast to the predictions of typical nebular models, to which the term {\it flash heating} is often applied. The peak surface temperature on the SPP, T(0), can, under no circumstances exceed T$_s$, the temperature of the lava on the LDP responsible for the heating. Again, as noted in Paper I, predicted peak temperatures and cooling rates are compatible with the severe constraints imposed by observed chondrule textures and experimental petrology \citep{Hewins1990,Desch2012}.

\begin{figure}
\label{fig1}
\plotone {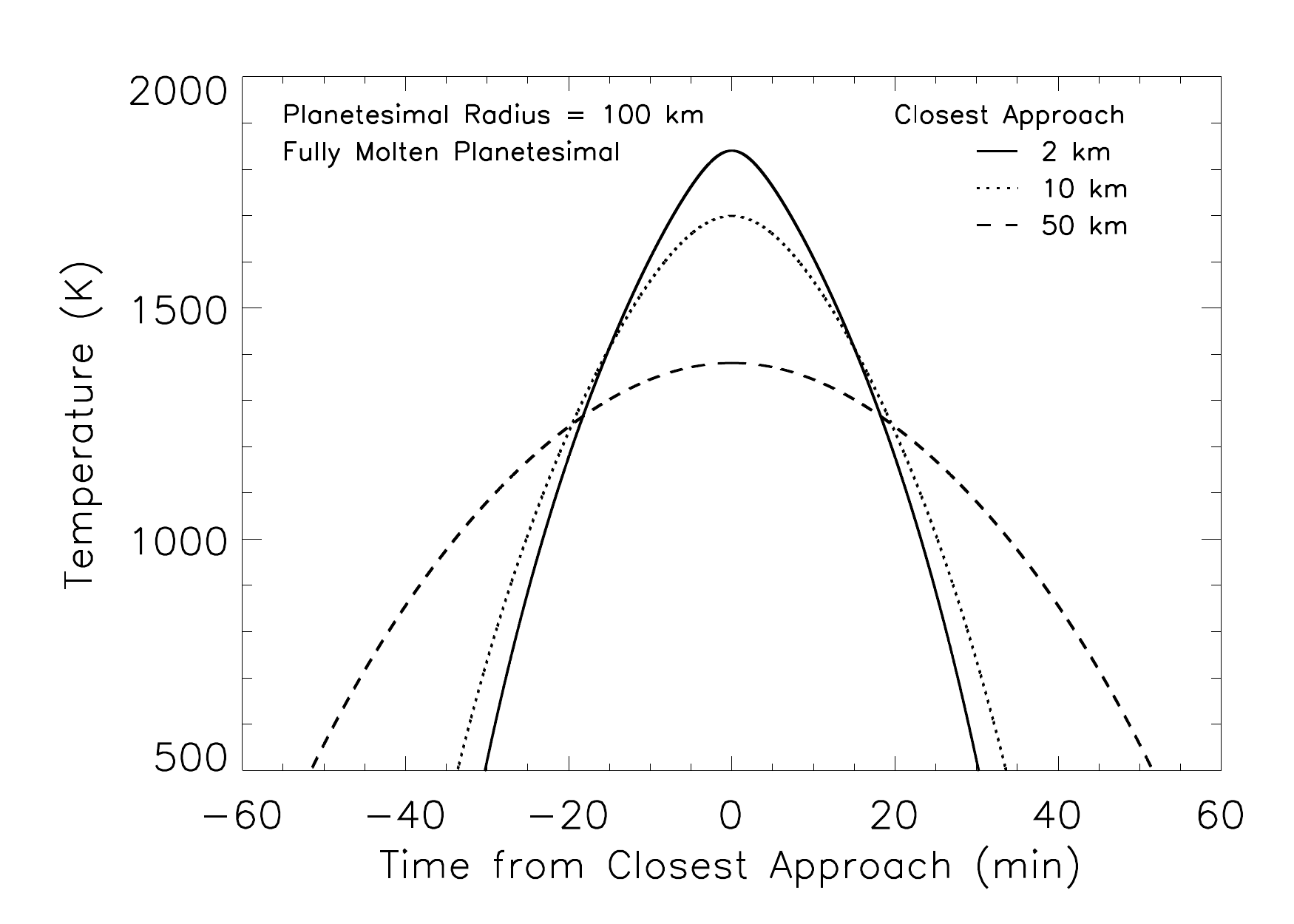}
\caption{SPP surface temperature versus time for flybys of a fully molten LDP of radius 100 km with $\rho$ = 3 gm cm$^{-3}$, $T_s$ = 2000 K, A = 0.04 and $\phi_b(0) = 0.93$.}
\end{figure}

\subsection{The Case of Lava Oceans of More Limited Extent}

The calculation of $\Omega$, the solid angle subtended by the hot lava on the LDP as seen from the surface of the SPP, is not entirely trivial for lava oceans of limited extent. 
In general, one may write
\begin{equation} 
\Omega = \Omega_p A
\end{equation} 
where A is the ratio of the projected area of the hot lava on the sky to the projected area of the whole LDP, as seen from the surface of the SPP. As an illustration of the effect of ocean size on the predicted thermal histories we consider a spherical LDP with a single, circular lava ocean. For ease of calculation we assume that the spot center is located on the equator of the LDP, which we also take to be the orbital plane of the SPP. The center of the spot is further fixed at the sub-surface point during periapsis, i.e. the SPP is constrained to fly directly over the center of the hot spot. Of course these assumptions could be relaxed in order to fit a particular heating/cooling curve in detail but our intention here is just to illustrate the effect.

The situation of a circular ``spot" of different temperature from its surroundings on a spherical object has been studied extensively by astronomers to model the brightness variations of stars with (usually cool) spots. Elegant treatments are provided by \citet{Budding1977} and \citet{Dorren1987}. Neglecting the effect of limb darkening, which is important in gaseous stellar atmospheres, but not here, one can see from either of the cited sources that, viewed from a great distance, 
\begin{equation} 
A= sin^2\ \gamma\ cos\ \alpha
\end{equation} 
where $\gamma$ is the angular radius of the spot and $\alpha$ is the angle between the line connecting the spot center to the LDP center of mass and the line of sight from the heating point on the surface of the SPP to the LDP center of mass. This applies only to the case that the spot is fully in view from the SPP on the visible hemisphere of the LDP. Once it is beyond the horizon from the viewing spot, the situation is more complicated. 

While the spotted star examples provide nice guidance to the current problem, the analytical solutions derived cannot be used directly here because we do not view the hot lava on the LDP from a great distance. Therefore, we do not see an entire hemisphere of the LDP at one time, but only a more limited portion of its surface, set by the distance from which we view. It may be possible to extend the formalism of the stellar case to the more general problem encountered here, but that is beyond the scope or need of this project. Instead, we have used a simple numerical integration over the hot spot area to calculate A. The custom code that we developed to do this was checked by comparison with analytical results derived in the stellar cases, and in other simple cases. 

Figs. 5 and 6 exhibit our results for two cases considered, $\gamma = 60\degr$, corresponding to a 25\% coverage of the LDP surface and $\gamma = 30\degr$, corresponding to a 7\% coverage. It is clear from Figs 4, 5 and 6 that the effect of constraining the hot lava to a smaller portion of the LDP's surface is to shorten the time that it spends at its highest temperature. The value of the peak surface temperature on the SPP does not change much because, at closest approach, the irradiated surface sees a hot lava ocean covering nearly or entirely the same field of view (i.e. horizon to horizon, as viewed from the SPP) that it would see regardless of the true extent of the hot lava. Obviously, non-central crossings of lava oceans would have similar effects -- shortening the high temperature exposure time but not changing the maximum temperature unless the spot were not crossed at all or barely crossed. Again, we emphasize that the precise contour, especially the rapid rate of decline is, in part, due to the unrealistic assumption of a single temperature to characterize the lava ocean. It is very likely that the surface around the lava ocean would be quite hot as well, even if somewhat capped over. We leave simulation of that case to a future study. 

\begin{figure}
\label{fig2}
\plotone {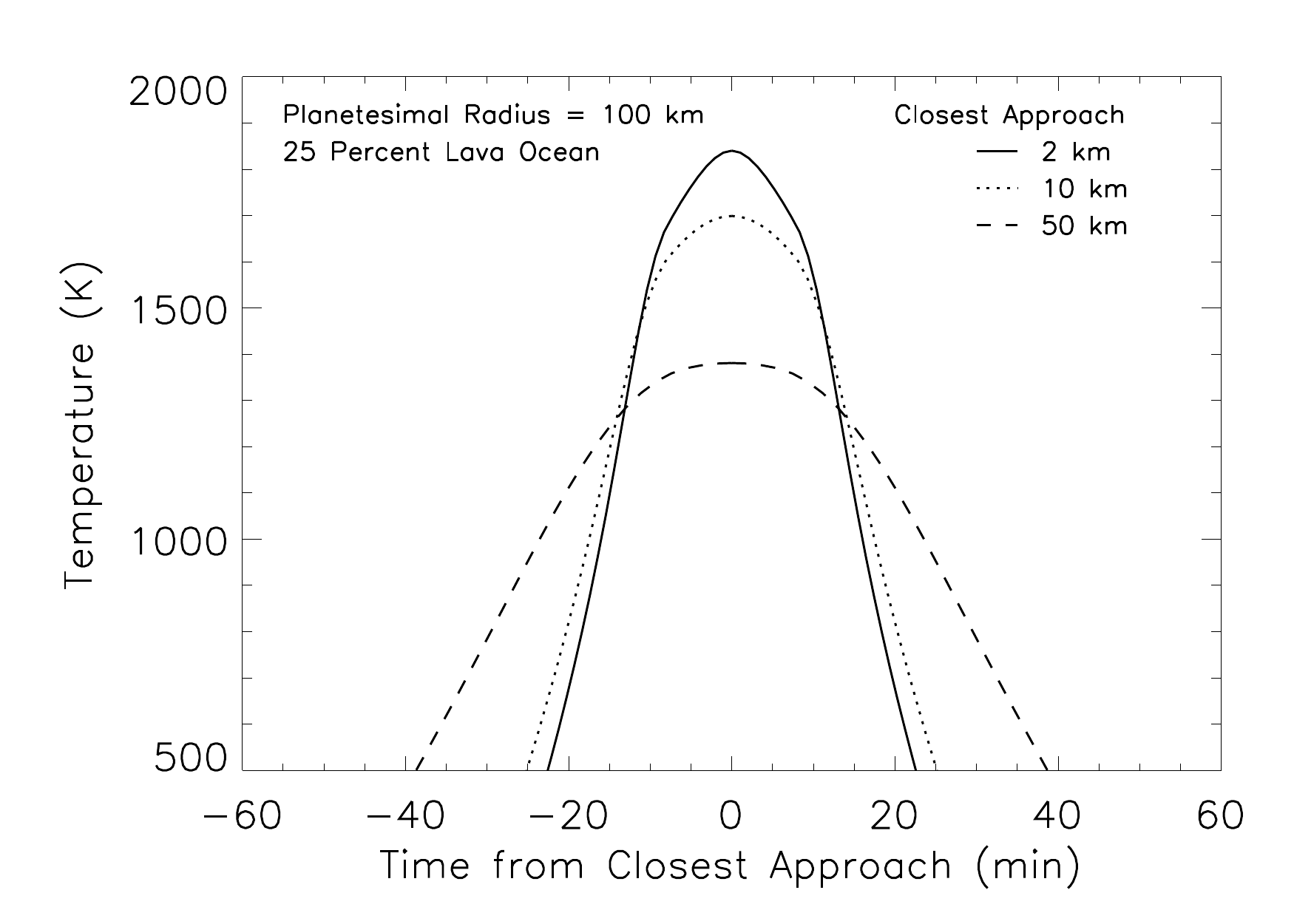}
\caption{SPP surface temperature versus time for flybys of an LDP of radius 100 km with $\rho$ = 3 gm cm$^{-3}$, $T_s$ = 2000 K, A = 0.04 and $\phi_b(0) = 0.93$.}
\end{figure}

\begin{figure}
\label{fig3}
\plotone {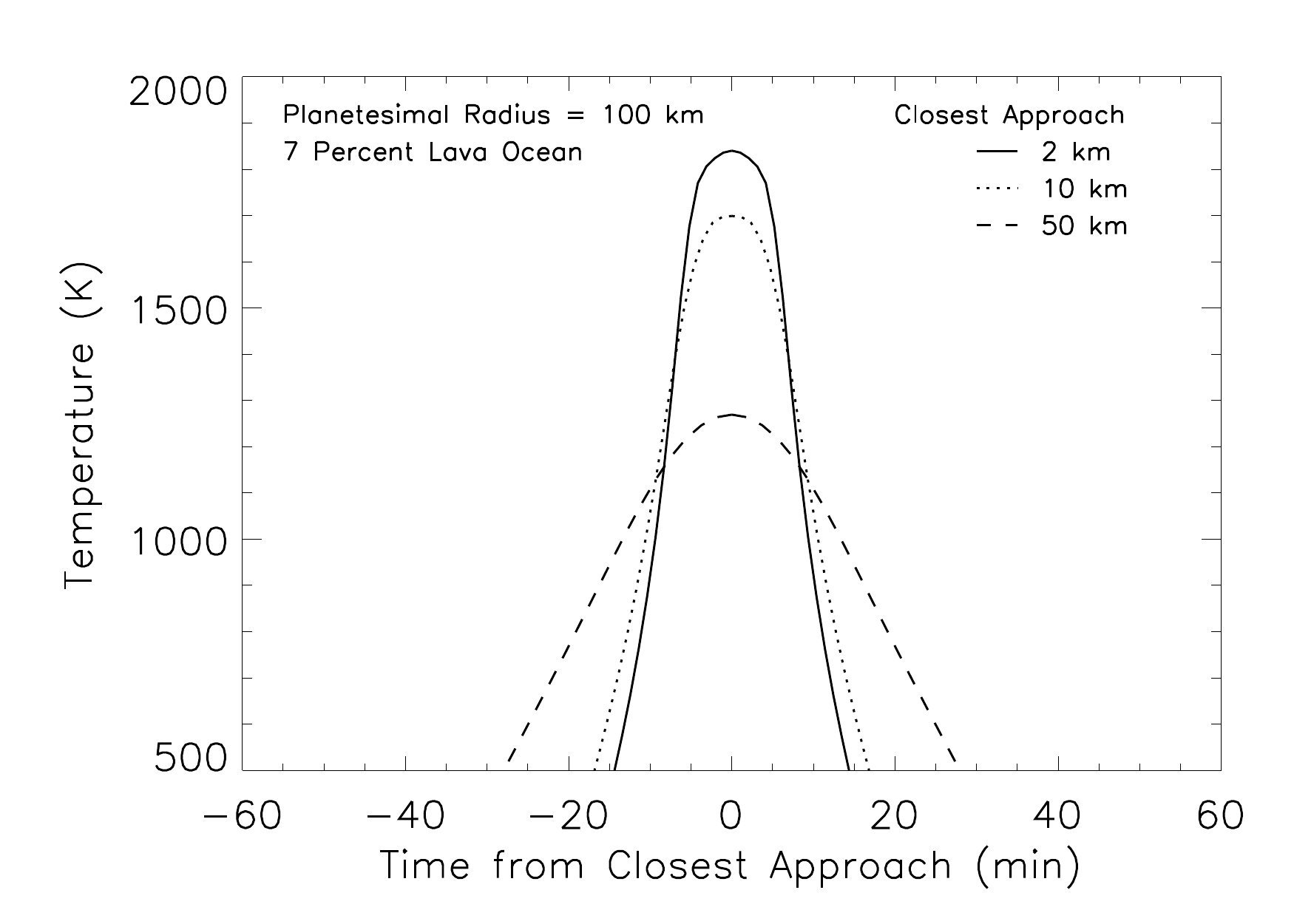}
\caption{SPP surface temperature versus time for flybys of an LDP of radius 100 km with $\rho$ = 3 gm cm$^{-3}$, $T_s$ = 2000 K, A = 0.04 and $\phi_b(0) = 0.93$.}
\end{figure}

\subsection{The Cases of Circular, Elliptical and Hyperbolic Orbits}

In Paper 1 and the examples above we have adopted a parabolic orbit. For completeness, and because it may actually occur in nature, we consider the cases of hyperbolic and bound (circular and elliptical) orbits. Not surprisingly, hyperbolic orbits shorten the time during which an object would be heated to interesting temperatures, while elliptical orbits lengthen it. The limiting case of the circular orbit provides the slowest heating/cooling rates that can be achieved without invoking phenomena such as rotation of the planetesimals or multi-component temperature models for the lava, both of which are reasonable phenomena to explore in future investigations. As noted in Section 2, It is possible that, during the t = 1--4 Myr epoch of chondrule formation, SPPs were common as ring particles or moonlets orbiting LDPs, either because they were captured into such orbits or were born into them. 

In Fig 7 we show three examples of flyby heating of SPPs on circular orbits of 2, 10 and 50 km height above the surface. We have chosen a lava ocean covering 25\% of the LDP surface for this illustration. The flat tops of the curves indicate the times during which, viewed from the surface of the SPP, one would see only hot lava on the LDP, covering almost an entire hemisphere. The sharp edges to these heating curves, again, are unrealistic and due to the choice of a single component temperature model for the spot. A more realistic distribution of surface temperature on the LDP would produce less extreme cooling rates. If the LDP had multiple eruption sites at one time, then the SPP could experience an intricate heating and cooling pattern. Low level heating episodes, insufficient for chondrule formation but perhaps sufficient for metamorphism and other phenomena, may occur, as well. Elliptical orbits naturally produce heating and cooling results intermediate between the circular and parabolic orbit cases.

\begin{figure}
\label{fig3a}
\plotone {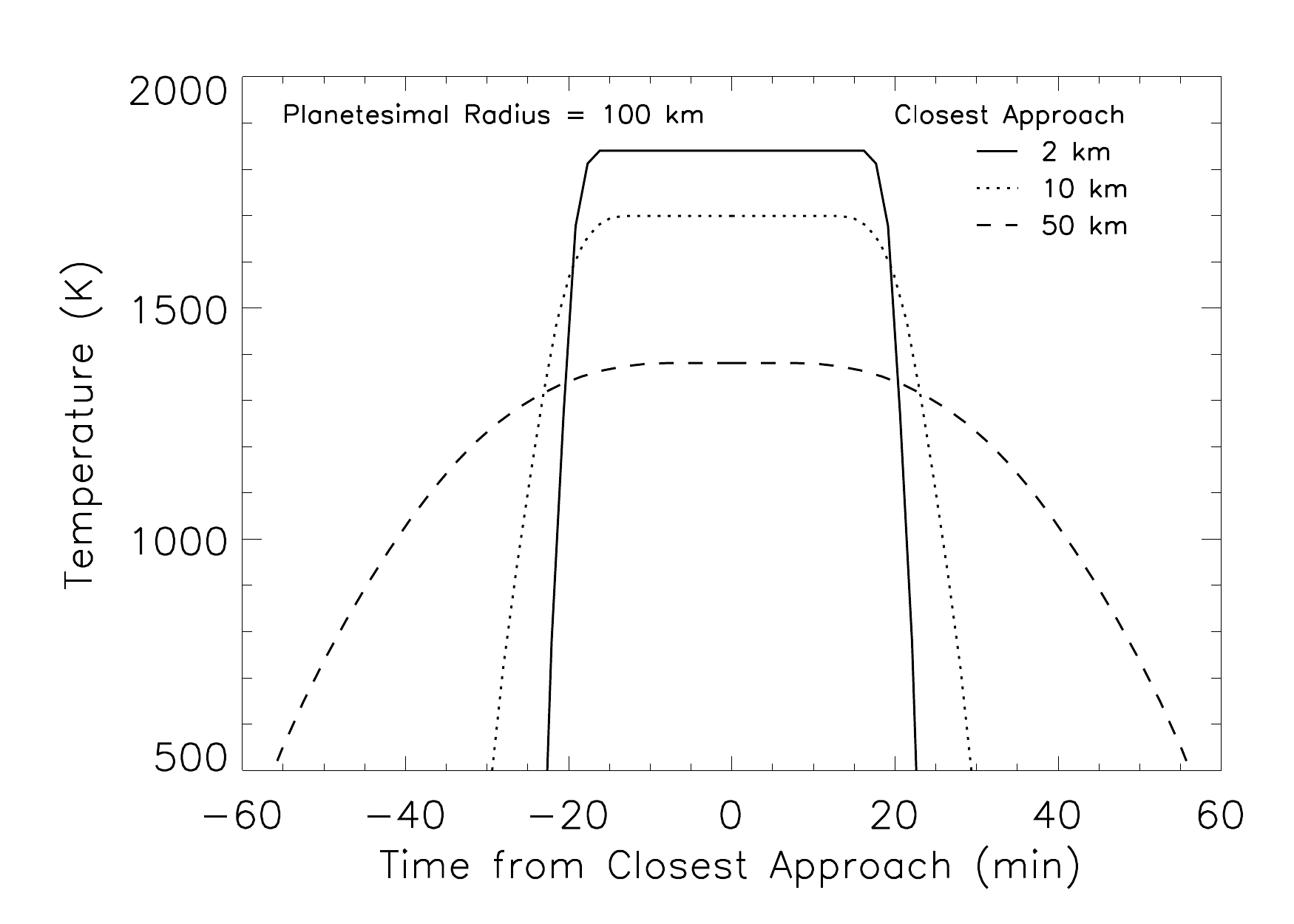}
\caption{ SPP surface temperature versus time for flybys of an LDP of radius 100 km with $\rho$ = 3 gm cm$^{-3}$, $T_s$ = 2000 K, A = 0.04 and $\phi_b(0) = 0.93$. A lava ocean is assumed to cover 25\% of the surface of the LDP. The orbit of the SPP in this case is circular at the given height above the LDP's surface and passes directly over the center of the circular lava ocean.}
\end{figure}

\subsection{The Temperature Profile with Depth in the SPP}

How deeply into the surface layers of the SPP can one expect the temperature to be raised by a sufficient amount to form chondrules? In terms of optical depth, the answer is given by Equation 8, which we may rewrite in terms of the surface temperature, T(0), as
\begin{equation} 
T(\tau) = T(0) (\phi_b(\tau))^{1 \over 4}.
\end{equation} 
$\phi_b(\tau)$, in turn, depends on the full optical thickness of the medium, $\tau_D$. The T-$\tau$ relationship for three choices of $\tau_D$ is shown in Figure 8, based on the calculations of \citet{Heaslet1965}. 

\begin{figure}
\label{fig7}
\plotone {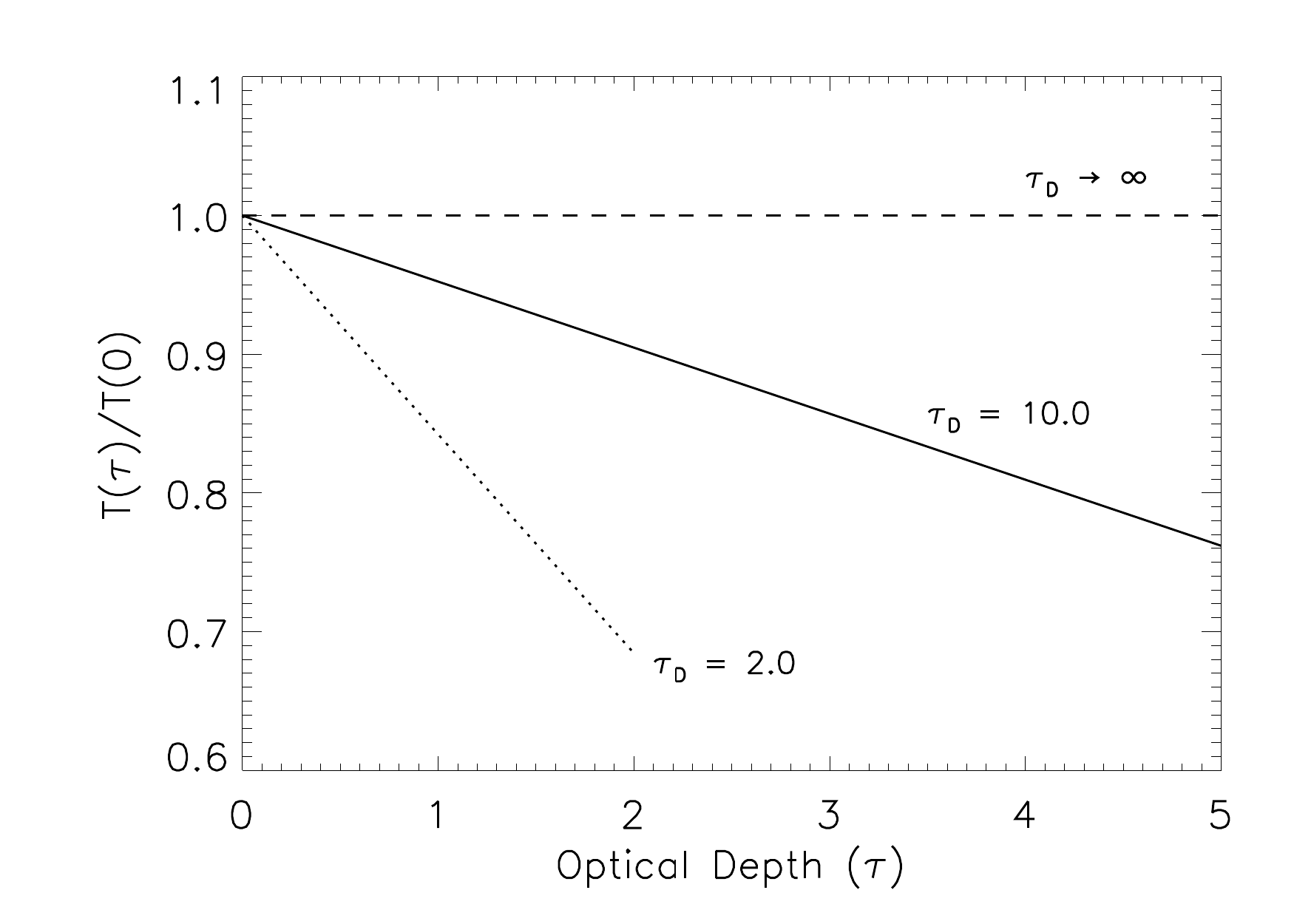}
\caption{Temperature with optical depth into the SPP normalized to its surface temperature, T(0), for three values of $\tau_D$, the full optical depth to which thermodynamic equilibrium is achieved.} 
\end{figure}

To translate optical depth into physical depth (s) one needs to specify the opacity ($\kappa$) and effective density ($\rho_{s}$) of the surface layers of the SPP, since
\begin{equation} 
\tau = \kappa \rho_{s} s.
\end{equation} 
$\kappa$ depends on the composition and size distribution of the grains. Here we consider spherical amorphous silicate particles ({\it astrosilicates}), whose optical properties have been well studied by \citet{Draine1984} and \citet {Laor1993}. See \citet{Eisenhour1995} for a discussion of the optical properties of other potential chondrule precursors including Fe metal and forsterite olivine grains. \citet{Draine1984} and \citet {Laor1993} have used Mie scattering theory to calculate the absorption efficiency, $Q_{abs}$, which is the ratio of a particle's effective absorbing cross-section to its geometric cross-section, for spherical particles with a variety of sizes. In Fig. 9, the variation of Q$_{abs}$ with wavelength for these astrosilicates is shown for three representative particle sizes from their tabulation at URL: https://www.astro.princeton.edu/$\sim$draine/dust/dust.diel.html. This is compared with the distribution of intensity for a Planck function at T = 2000 K. It is clear from the figure that particles of radius 10 $\mu$m (or larger) have values of Q$_{abs}$ close to unity, but as grain size decreases, the particles become significantly less efficient at absorbing energy from the ambient radiation field. This depends somewhat on the composition of the grains, but primarily on their size. 

In this exploratory study, we model the irradiated medium as an aggregate of uniform spherical (pre-chondrule) grains of radius, a, and grain density, $\rho_g$, embedded in a lattice of fine-grained material with a closed pore structure (pre-matrix). CAI's and pre-solar grains are included in the mix but have no effect on the opacity, as they are are minor components. In this model, the opacity derives entirely from the pre-chondrule grains, which are assumed to compose half the mass, a typical value for chondrites. For a $\gg$1 $\mu$m, it follows that
\begin{equation} 
\kappa = {3 \over {2 a \rho_g}},
\end{equation} 
where we have used the well-known result from Mie scattering theory that Q$_{ext}$ $\approx$ 2 Q$_{abs}$ for particles large compared to the wavelength of irradiation. The factor of 2 arises from the scattering portion of the opacity caused by diffraction at the edges of the particles. The mean free path of a photon in this medium (s$_{mfp}$) then follows from Equation 13, by setting $\tau$ = 1 and $\rho_s = {\rho_m \over 2}$, where $\rho_m$ is the density of the SPP,
\begin{equation} 
s_{mfp} = {4 a \rho_g \over {3 \rho_m}}.
\end{equation}
If we choose $\rho_g$ = 3 gm cm$^{-3}$ and a = 0.05 cm, representative of actual chondrules, then $\kappa$ = 10 cm$^2$ gm$^{-1}$. If we further adopt 
$\rho_m$ = 10$^{-4}$ gm cm$^{-3}$ based on the constraints of \citet{Alexander2008}, then s$_{mfp}$ = 20 m, identical to what we suggested in Paper I. 

It remains to determine what value of $\tau_D$, the full optical thickness of the region in thermal equilibrium, is appropriate for this model. First, we show that the individual pre-chondrule grains will come into thermal equilibrium with the radiation field very quickly -- a matter of seconds or less. Following \citet{Draine2011} we may write that the radiative heating rate is
$$\big( {dE \over dt} \big) = {<Q_{abs}> _{rad} \pi a^2 u_{rad} c} $$
where $dE$ is the amount of heat energy added to the grain in time $dt$,  $<Q_{abs}> _{rad}$ is the wavelength averaged absorption efficiency weighted by the relevant distribution of radiative intensity, a is the particle radius, $u_{rad}$ is the energy density of the radiant energy field and c is the speed of light. We have computed $<Q_{abs}> _{rad}$ for T = 2000 K and display the results in Figure 10. For large grains $<Q_{abs}> _{rad} $ $\sim$ 1, while for a grain with a = 1 $\mu$m,  $<Q_{abs}> _{rad} $= 0.63 (for T = 2000 K), and for a=0.1~$\mu$m, $<Q_{abs}> _{rad} = 0.03$. 

To compute the temperature change, $\Delta$T, associated with an energy input, $\Delta$E, one requires the specific heat (heat capacity per unit mass) of the material, C, and the mass of the grain being heated, m$_g$, in which case,
\begin{equation} 
\Delta E = C m_g \Delta T.
\end{equation} 
For spherical grains of fixed density, $\rho_g$, the mass of a grain is $m_g = {{4 \over 3} \pi a^3 \rho_g}$ and for a black body radiation field of temperature, T, the radiant energy density is 
\begin{equation}  
u_{rad} = {4 \sigma T^4 \over c}. 
\end{equation}
Combining expressions and simplifying we have
\begin{equation}  
\Delta t = {{\rho_g C a} \over {3 <Q_{abs}>_{T} \sigma {T}^4}}\Delta T  
\end{equation} 
where $\Delta t$, is the time interval necessary to raise a grain by $\Delta T$ to its steady state temperature in a given radiation field, characterized by T.   Adopting representative values for silicate grains of $\rho_g$ = 3 gm cm$^{-3}$ and C = 0.84 J gm$^{-1}$ K$^{-1}$, we can calculate $\Delta t$ as a function of grain size for a given radiation field. Fig. 10 shows the heating time necessary to raise the temperature of a grain by 100 K, if it is embedded in a black body radiation field characterized by T = 1700K, as a function of grain size. Clearly, grains up to 1 mm would reach the equilibrium temperature of their surroundings very quickly -- a matter of seconds. Chondrule-sized objects would, therefore, be expected to heat up and maintain a temperature close to thermal equilibrium throughout the flybys. 
\begin{figure}
\label{fig4}
\plotone {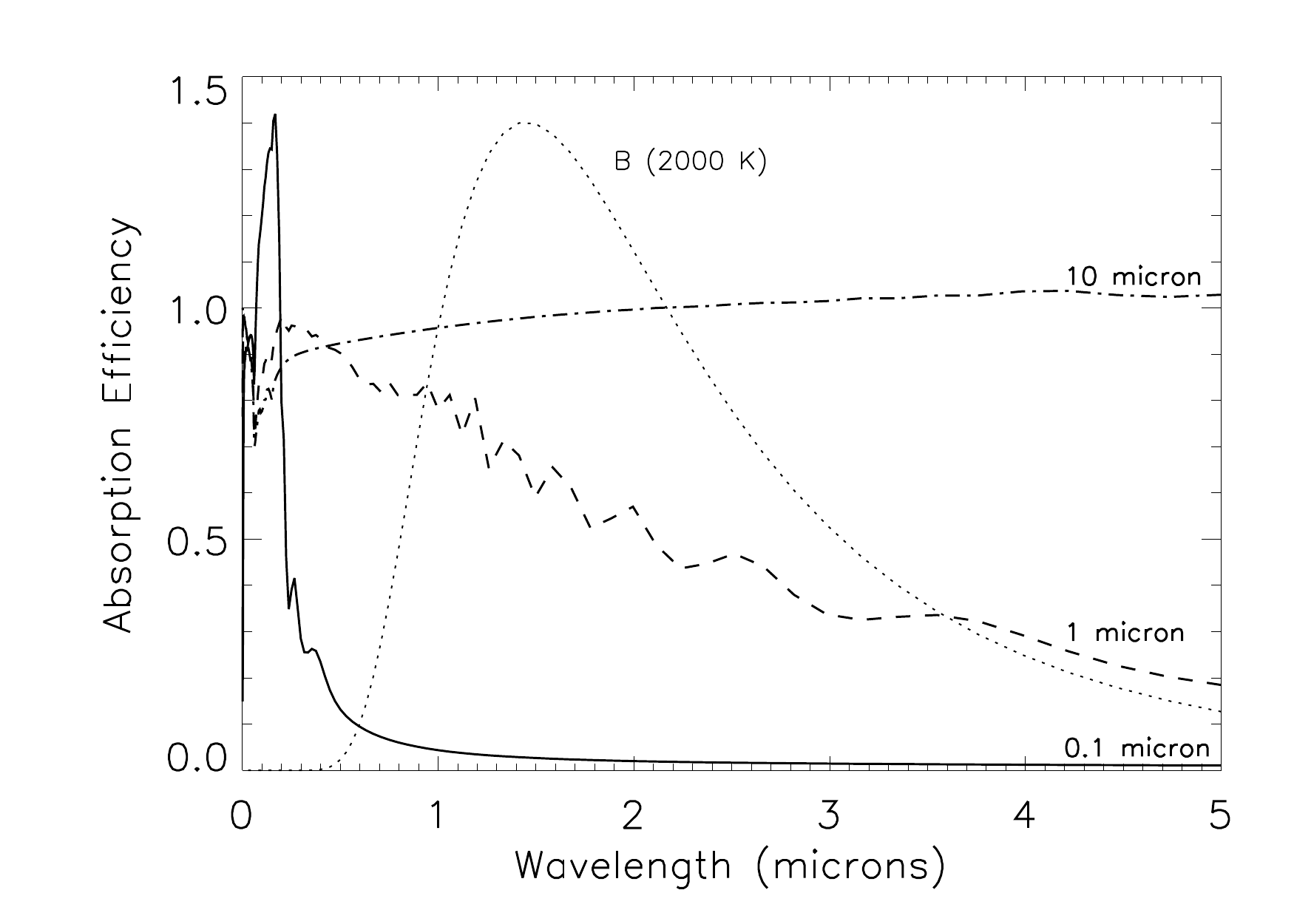}
\caption{Absorption Efficiency (Q$_{abs}$) for {\it astrosilicates}, amorphous silicates, from URL: https://www.astro.princeton.edu/$\sim$draine/dust/dust.diel.html \citep{Draine1984, Laor1993} as a function of wavelength for three different spherical particles of given radius. Also shown is the scaled Planck function for a temperature of 2000 K. Sub-micron sized particles are inefficient at absorbing energy from a black body with a temperature near 2000 K. Grains larger than 10 $\mu$m have absorption cross-sections approximately equal to their geometric cross-sections, Q$_{abs}$ $\sim$ 1.} 
\end{figure}

\begin{figure}
\label{fig5}
\plotone {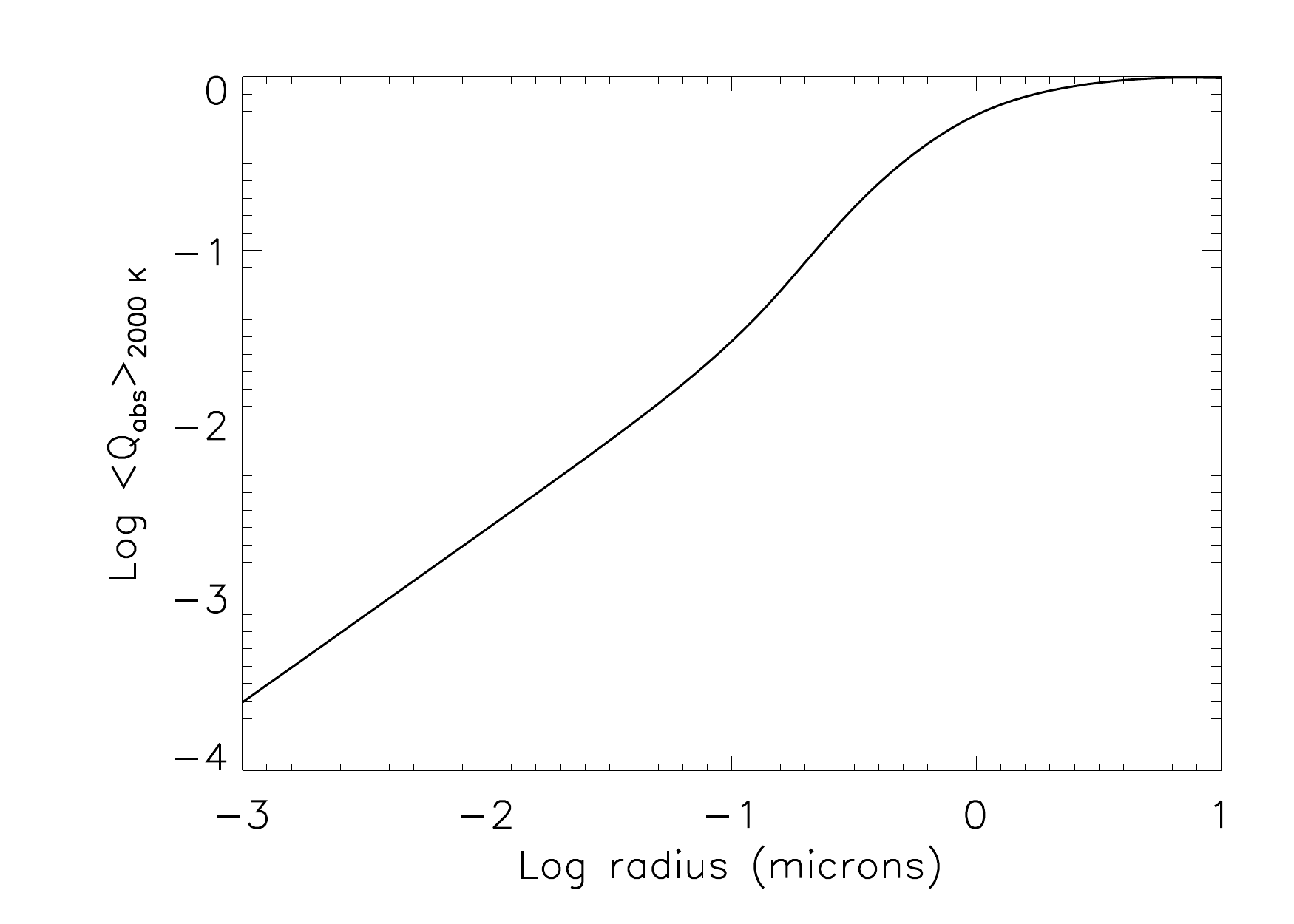}
\caption{Absorption Efficiency ($<$ Q$_{abs}>$) for {\it astrosilicates}, amorphous silicates, from URL: https://www.astro.princeton.edu/~draine/dust/dust.diel.html \citep{Draine1984, Laor1993} as a function of grain radius for a blackbody temperature of the radiation field of $T_s$ = 2000 K.} 
\end{figure}

\begin{figure}
\label{fig8}
\plotone {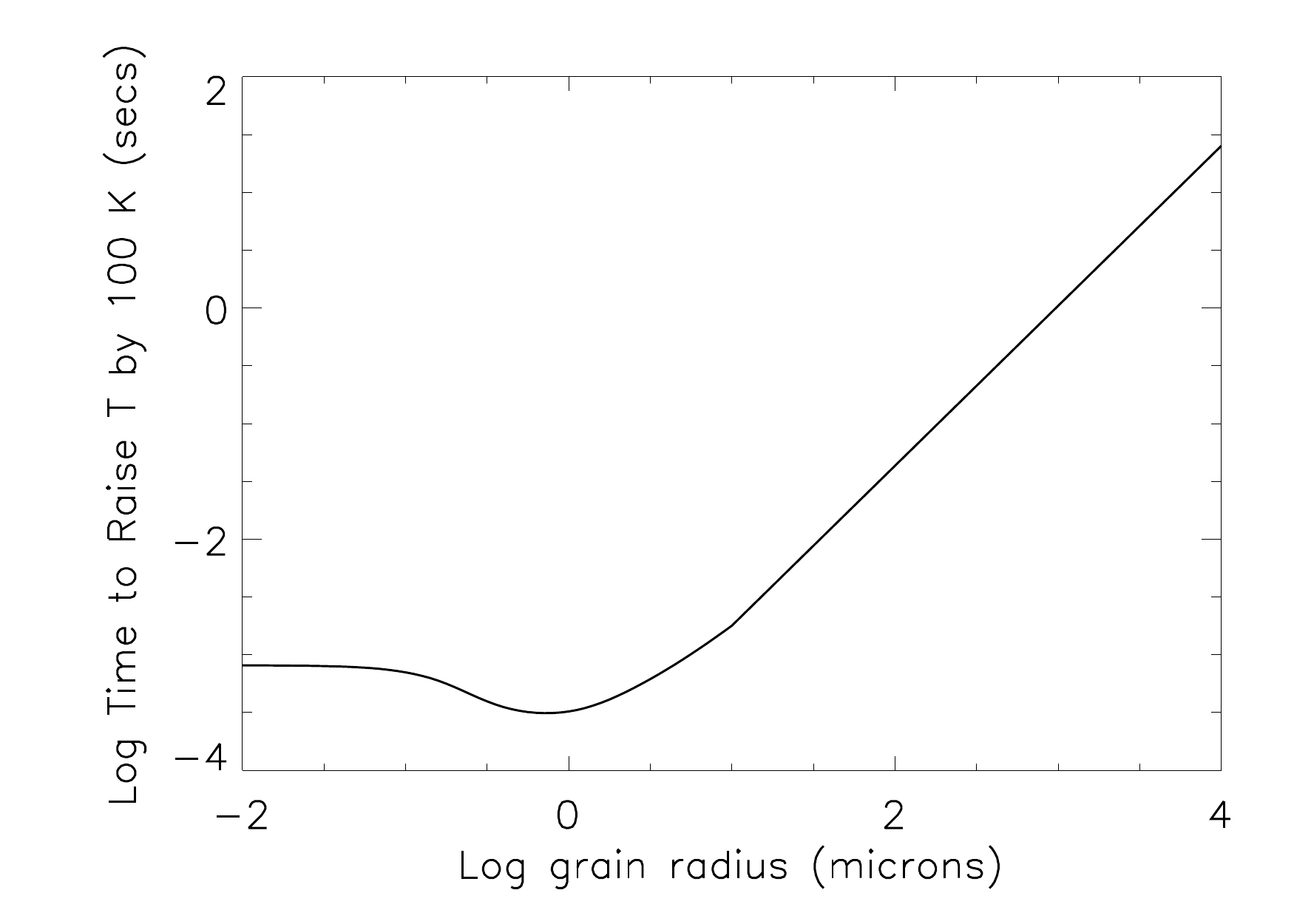}
\caption{Time for a spherical grain to reach equilibrium temperature if exposed to a radiation field from a 1700 K black body as a function of grain radius, a. A heat capacity of C = 0.84 J gm$^{-1}$ K$^{-1}$ and grain density of $\rho_g$ = 3 gm cm$^{-3}$ are assumed, characteristic of sand particles.}
\end{figure}

There is, of course, only a finite amount of energy available to heat grains during a flyby, and this will set a limit on how large $\tau_D$, the full optical depth of the medium that is in thermal equilibrium can become. As $\tau_D$ increases, the net flux of energy through the medium decreases because T(0) approaches T$_1$. Again, following Chapter 11 of \citet{howell2016}, we define the dimensionless energy flux, $\Psi_b$, as the ratio of the actual energy flux, which is constant through the medium under the assumption of radiative equilibrium, to the flux incident on the surface, $\sigma T_1^4$. \citet{Heaslet1965} provide an approximation to $\Psi_b(\tau_D)$ as
\begin{equation}  
\Psi_b(\tau_D) = {\big({4 \over 3}\big) \big({1 \over {1.42089 + \tau_D}}\big)} 
\end{equation}
which we adopt here. For any given $\tau_D$ we can, therefore, determine the energy flux through the medium and, over the flyby, determine the total amount of energy available for heating the layers. 

The procedure adopted for determining $\tau_D$ was to make an initial guess of the value, determine $\Psi_b$ from Equation 19, and compare the integrated flux available for heating the layer to a depth, D, with the amount of energy required to raise the temperature of the grains to their equilibrium value. Equality between the flux provided and the flux required was achieved for a value of $\tau_D = 24$ with the parameters in Table 1. Combining this result with s$_{mfp}$ from above, we find D $\approx$ 500 m for the depth to which equilibrium heating would be achieved given the choice of parameters. Of course, the temperature at D, in this example, will be below that required to form chondrules; one might expect that to occur to perhaps a depth of 100 -- 250 m, depending on how close the flyby is. Also note that the length scale depends on $\rho_m^{-1}$ and this is probably the least constrained parameter in the model. For example, if $\rho_m$ is increased to 10$^{-3}$ gm cm$^{-3}$ then heating will only extend about 50 m into the object. 

A distinctive feature of the flyby model is that the mass of chondrites formed by this mechanism is limited, because sufficient heat to form chondrules can only penetrate into the surface of the SPP by a limited amount during the limited time of the flyby. Roughly, one may expect the scale length for chondrule formation to be about 10~s$_{mfp}$. Employing Equation 15 with a = 0.05 cm and $\rho_g$ = 3 gm cm$^{-3}$ allows us to translate this to a characteristic linear scale for chondrite formation. Identifying that scale as the initial diameter of the largest monolithic chondrite that could form and using the initial density of the material, $\rho_m$ to translate that to a mass yields  
\begin{equation}  
M_{chondrite} \le {4.2\ ({  \rho_m \over 1\ {\rm gm\ cm^{-3}} })^{-2}}\ {\rm gm},
\end{equation}
where $\rho_m$ is the initial surface layer density of the SPP in gm cm$^{-3}$. For $\rho_m = 10^{-4}$ gm cm$^{-3}$ we find a maximum mass of about 420 t, which is more than two orders of magnitude larger than the largest known chondrite. As we discuss below, this model predicts that larger meteoroids would not be monolithic, but assemblages of individual stones smaller than 420 t, and probably much smaller, since we have used extreme values here.  

\section{Laboratory Simulation of Chondrule Textures}

As a demonstration of the ability of the flyby model to replicate important chondrule properties, we show the results of a small set (12) of laboratory experiments to simulate Type I (FeO-poor) porphyritic olivine (PO) chondrules.  The FeO-poor PO chondrules are the most voluminous type of chondrule in carbonaceous, ordinary, and enstatite chondrites \citep{Jones2012}.  A suite of minerals, chosen based on previous experimental studies of chondrule synthesis by \citet{Radomsky1990}, \citet{Connolly1998}, \citet{Whattam2009} and \citet{Villeneuve2015} were subjected to a published thermal trajectory of a 10 km planetesimal, with 100\% lava coverage, and a 4 km closest approach (Fig. 2 of Paper I). The range of mineral compositions is shown in Fig. S1 and detailed in the Supplementary Material (SM). The match between the experiment's actual heating and cooling history and the published thermal trajectory is shown in Fig. S2.  Experimental methods can be found in the SM.  

In Figure 12, experiment PB-48c is compared with a Type I PO chondrule from Semarkona, showing the similarity in olivine grain size and texture.  Figure 13 shows the Mg\# (Mg/Mg+Fe (atomic)) of olivine in the resulting experimental charge and the chemical composition of the experimental glass, normalized to the average chondrule glass of Type I PO and porphyritic olivine pyroxene (POP) chondrules from Semarkona \citep{Alexander2008}. Type I PO chondrule olivine is typically closer to a Mg\# of ~99 \citep{Jones1989}, and this composition is achieved in half of the experiments (Figs. S3-S5, S7-S8, S13).  The experimental glasses are similar to chondrule glasses for most major elements except Na$_2$O and FeO. The Na$_2$O is being volatilized from our charge, and we were not trying to control for this.  The resulting high FeO is an experimental byproduct of our precursor materials, which are terrestrial mantle minerals, and consist of oxidized iron in silicates.  Half of the experiments are within a factor of 3 of FeO of chondrule glasses (Figs. S3-S8).  
	
The experiments demonstrate a reasonable reproduction of Type I FeO-poor PO chondrule texture, as well as major-element chemistry of olivine and glasses, using a published heating and cooling rate of the flyby model.  This demonstrates, if chondrule precursors were similar to what we have used, that the flyby model can satisfy the first order thermal constraints imposed by chondrule texture and major element chemistry \citep{Connolly2016}.  Future experiments will aim to reproduce chondrule zoning of Ti and also oxygen isotopes in chondrule olivines \citep{Marrocchi2018}, as this should be possible with our current experimental system, and would allow for further assessment of this model's potential in reproducing observed characteristics of chondrules. Mineralogical zonation is a common feature of Type I chondrules \citep{Friend2016} and has been suggested to result from interaction of molten chondrules with a Si-bearing gas \citep{Tissandier2002}.  Higher partial pressures of Si and Na are expected in the SPP during heating and could explain mineralogical zonation of chondrules.  Experimental reproduction of mineralogical zoning due to high partial pressures of Si are not possible with our current experimental setup, but could be an exciting new avenue of future research in chondrule synthesis.    

\begin{figure}
\label{fig12}
\epsscale{0.9}
\plotone {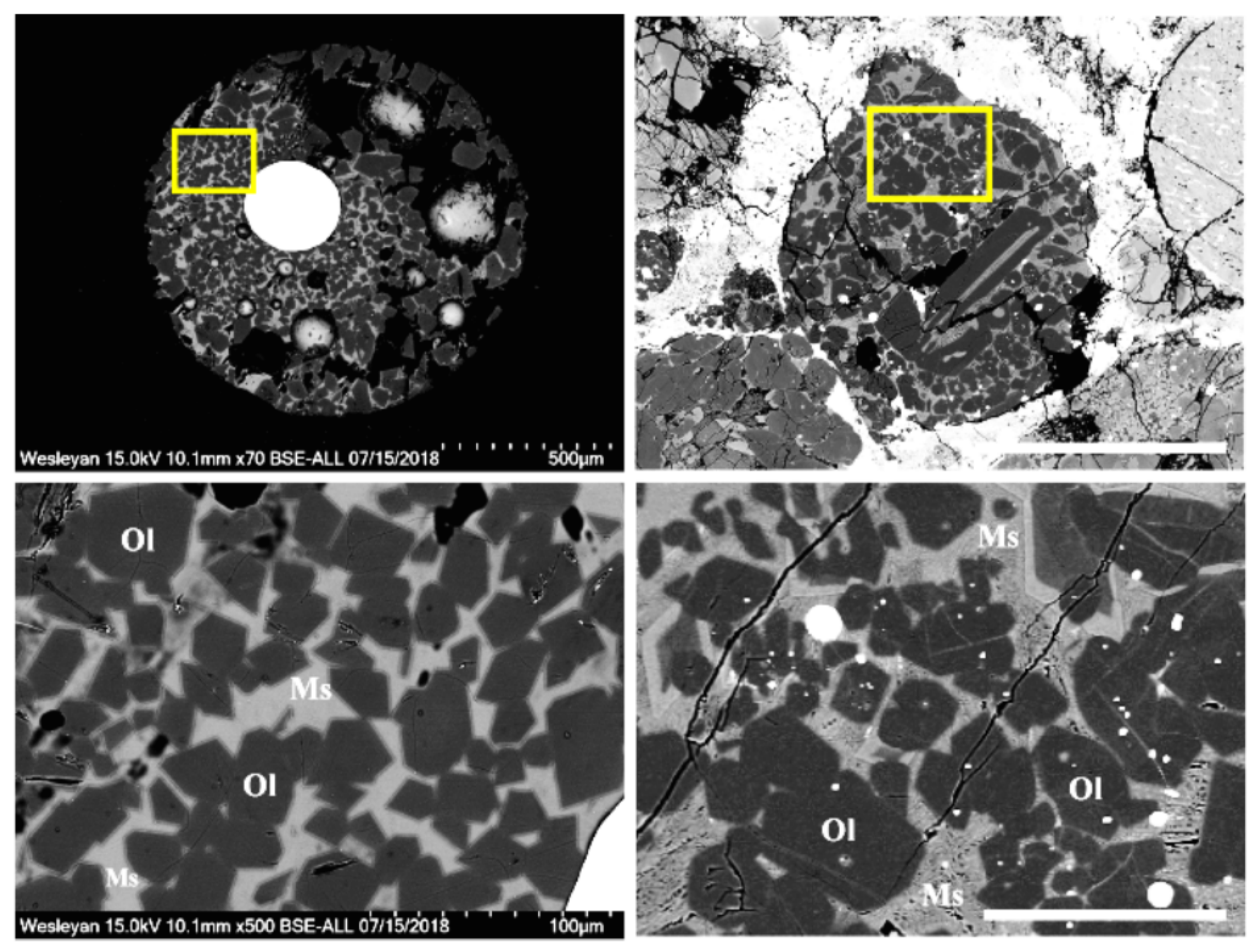}
\caption{(Top Left).  Backscatter Electron Image (BEI) of Experiment PB-48c.  The experimental charge consists of olivine, glass, vesicles, and Pt wire (the bright round object).  Scale bar is 500  $\mu$m. Yellow square denotes area detailed below.  (Bottom Left) Higher magnification BEI of olivine (Ol) and glass (Ms).  Scale bar is 100  $\mu$m.  (Top Right) BEI of PO chondrule from Semarkona.  Scale bar is 500  $\mu$m.  Yellow square denotes area detailed below.  (Bottom Right) Higher magnification BEI of chondrule shown in Top Right image.  Generally similar size of chondrule olivine to that of experiment PB-48c can be observed.  Scale bar is 100 $\mu$m.}
\end{figure}

\begin{figure}
\label{fig13}
\epsscale{0.5}
\plotone {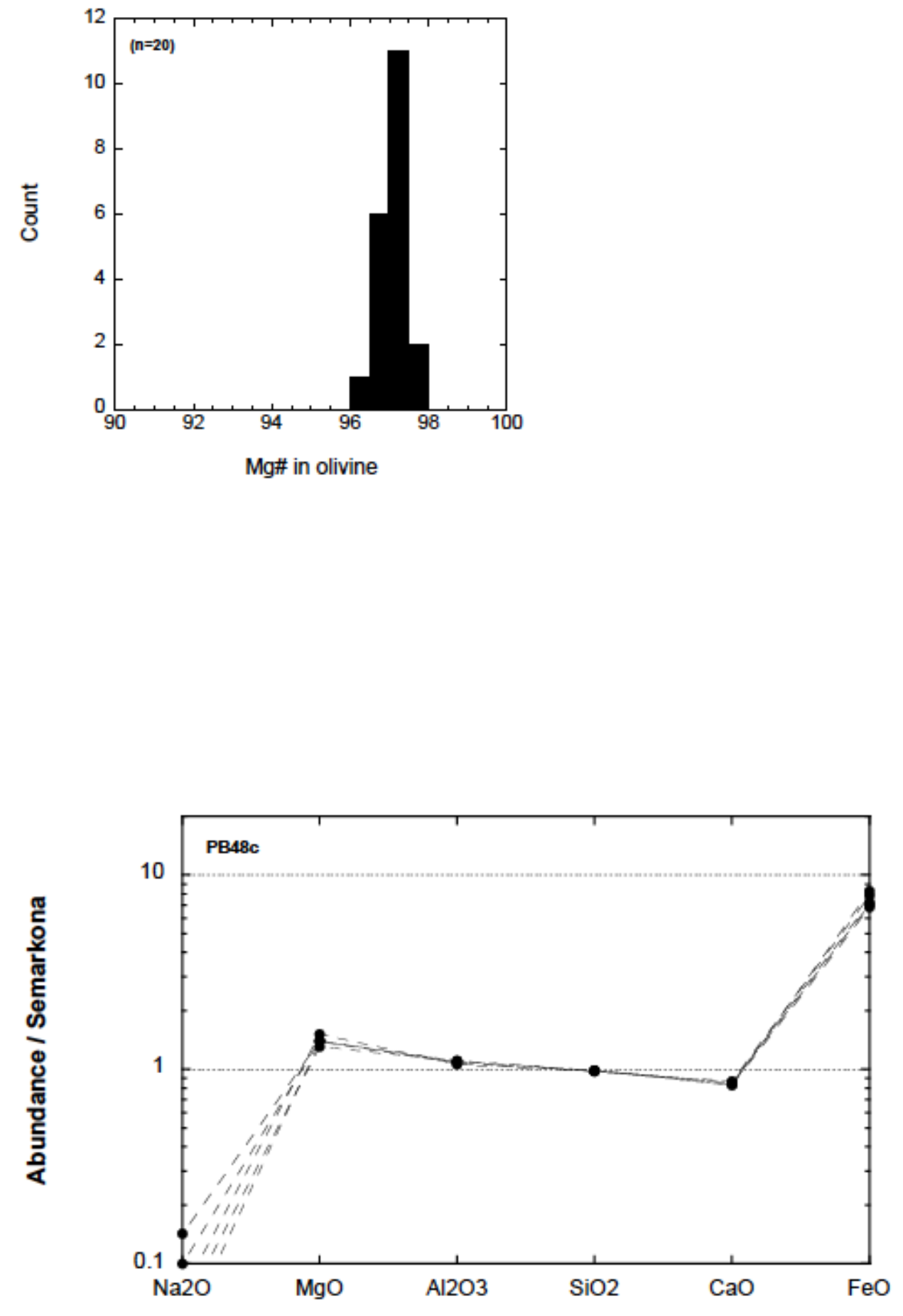}
\caption{(Top)  Mg\# of chondrule olivine grains in experiment PB-48c shown in Fig. 12.  (Bottom) Analyses of major element oxides of experimental glass in PB-48c, normalized to the average Type I PO and POP chondrule glass of Semarkona \citep{Alexander2008}.}
\end{figure}

\section{Discussion}

The model presented in this paper provides for a direct link between chondrule formation and chondrite lithification. In that sense it represents a major departure from the canonical view that chondrules formed first in space, then accreted to a parent body where they were lithified by processes unrelated to chondrule formation. As discussed above, a primary motivation for considering such a non-canonical model is complementarity, which provides strong evidence of a close link between chondule formation and chondrite lithification, and which no canonical model of chondrule formation currently addresses \citep{Lichtenberg2018}. Similarly, cluster chondrites and the size-class relationship of chondrules and chondrites suggest such a close link. In this section we elaborate on these and other implications of the flyby model, including the phenomena of metamorphic and aqueous alteration, the origin of CI chondrites (which have no chondrules), the abundance of chondrules in the early solar system, the evidence that chondrules and chondrites formed in weak magnetic fields, and the structure of meteoroids and primitive asteroids. Of particular interest is whether there is any evidence supporting the canonical view that renders the non-canonical model proposed here untenable.     

\subsection{Chondrite Lithification}

Heat, and some degree of pressure, are required to lithify chondrites sufficiently that they can survive the disruption of their parent body, a 10 -- 100 Myr transit from the asteroid belt to 1 AU, and subsequent passage through the Earth's atmosphere \citep{Consolmagno1999}. Due to the small sizes of asteroids, compaction by hydrostatic pressure is unlikely \citep{Consolmagno2002} and attention has focused on pressure pulses associated with impacts, as modeled, for example, by \citet{Beitz2016}. The difficulties with this approach are well illustrated in that paper and include the nature and retention of regolith on a parent body. Most chondrites do not show evidence of impact shocks in their history. Where evidence for shocks exists, it could be related to the impact that released the meteorite from its host body. If a series of relatively small compaction events associated with impact shocks is responsible for lithifying chondrules this would probably require hundreds of millions or billions of years and its relationship to the metamorphic classes of the OC's, which must be set up on a much shorter time frame, is unclear.

Elevated temperature facilitates lithification, reducing the pressure requirements substantially. A commercial technique based on this principle is known as hot isostatic pressing or HIP \citep{Atkinson2000}. It is employed commercially at typical gas pressures of 100 MPa to lithify fine powders by raising their temperatures to about 70\% of their melting point for times on the order of hours. \citet{Gail2015} have proposed that this mechanism is responsible for chondrite lithification on the H and L chondrite parent bodies, at the relatively low temperatures associated with chondrite metamorphism. The lower temperatures mean that longer times -- millions of years, instead of an hour -- would probably be required to acheive a substantial reduction in porosity. While this is a lengthy period of time, it is likely shorter than a cold pressing approach based on multiple impact shocks would require. There is no correlation of either shock class or metamorphic type with porosity, so neither lithification mechanism is directly supported by the meteorite record \citep{Britt2008}. An alternative possibility, proposed here, is that sintering at roughly atmospheric pressures ($\sim$0.1 MPa), but at the higher temperatures inferred for chondrule formation, will act to lithify the chondrites by the HIP processes on a much shorter time scale of minutes to an hour. 

We can roughly evaluate the pressure (P) and temperature (T) conditions in the chondrule and chondrite formation zone by assuming the ideal gas law 
\begin{equation}  
P = {{\rho_v \over {\mu m_H}} k T}
\end{equation}
where $\rho_v$ is the gas density, $\mu$ is the mean molecular weight in units of $m_H$, the mass of a H atom, and k is the Boltzmann constant. We may express $\rho_v$ as a fraction, x, of the surface density, $\rho_m$, of the SPP. Adopting T = 1700 K, $\mu = 16$, $\rho_m = 10^{-3}$ gm cm$^{-3}$ and x = 0.1 yields P = 0.09 MPa, close to 1 bar. The gravitational field of the SPP is negligible, but there will be a pressure gradient associated with the temperature gradient, which could act to provide some confinement of the material if the pore structure is open in parts. The expected temperature gradient in the chondrule formation zone follows from equations 12 and 13 and for any reasonable choice of parameters, we expect it to exceed 1 K m$^{-1}$. We conclude, therefore, that the overall conditions of pressure and temperature, as well as a mechanism of confinement, that apply to terrestrial implementations where the HIP process is effective in lithifying material could apply as well within the heated surface layers of an SPP during a flyby.

Circumstantially, several lines of evidence suggest that chondrites lithified roughly or precisely contemporaneously with the formation of chondrules. The strongest and best-known evidence for this comes from the phenomenon of complementarity, discussed in the introduction. In short, the observations demand that the chondrules and matrix formed from a closed system. If they did not form simultaneously during the same heating event, then one needs a mechanism to capture gas atoms released from the melting chondrules and condense them or otherwise trap them onto the fine-grained matrix material. The further removed in time and space that the chondrules are from the matrix, the more imposing the task. Complementarity is, therefore, strong evidence against hypotheses, such as the X-wind mechanism \citep{Shu2001}, in which chondrules form far from the site of chondrite formation. It also severely challenges all collisional models of chondrule formation, since there is no plausible way to store the atoms not incorporated into the chondrules for subsequent incorporation into the matrix \citep{Budde2016}. Even with nebular models it strains credulity to believe that atoms released during a chondrule formation episode can be kept for any lengthy period of time within well defined regions of space, such that grains of matrix with the right elemental compositions can condense and find their way back to the proper chondrules. \citet{Lichtenberg2018} point out that no current model of chondrule formation addresses complementarity.

A less well-known, but perhaps equally constraining piece of evidence on the relationship between chondrule and chondrite formation, comes from the study of so-called ``cluster chondrites" \citep{Metzler2012}. Cluster chondrites are portions of OCs in which chondrules of a variety of textures are so densely packed that they are no longer spheroids but have complex shapes obviously molded by the chondrules around them.The cluster chondrites are found in unequilibrated ordinary chondrules of all three groups (H, L, LL) and were found in 41\% of the investigated chondrites, comprising from 5\% - 90\% by volume, making them common.  The chondrules were clearly pliable objects at the time the chondrite was assembled and lithified. Since cooling rates for chondrules are of order hours (to days at most), these structures strongly argue that the formation of the chondrite occured within at most days (and probably minutes) of the formation of the chondrules. If that is true for the cluster chondrites, which have a sufficiently high density of chondrules to cause this shape modification it may be true for all chondrites, even if their chondrule density is too low to show a modified shape. Along these lines, we note that about 5\% of all chondrules are ``compound", showing that they were in close proximity to each other when they formed and that they lithified in such a way as to preserve that arrangement \citep{Ciesla2004}.

Another observed link between chondrules and chondrites that requires explanation is the correlation between chondrule size and chondrite group \citep{Friedrich2015}. If chondrules form independently in time and/or space from the chondrites they later populate, then one requires a size-sorting process to bring those of similar dimension together within a particular group of chondrites. Since the size differences are small (but significant), the size-sorting process must be very efficient, and so far there is no generally accepted mechanism for this. An alternative possibility, as suggested by \citet{Eisenhour1995}, is that chondrules formed by radiative heating from a source with a temperature measured in thousands of degrees. They argue that the distinctive sizes of chondrules within given chondrites might then be understood as arising from their optical properties, obviating the need for any sort of aerodynamical size-sorting. They further note that other distinctive properties of chondrules, including dusty rims, might also find their explanation in the fact that grains exposed to the same radiation field will reach much different peak temperatures due to the strong dependence of their optical properties on size and composition. For example, Fe metal particles absorb radiation much more efficiently in the infrared, than forsterite olivine, which is highly transparent at wavelengths near 1 $\mu$m. The radiative heating model proposed in this paper differs in some respects, particularly temperature, from that considered by \citet{Eisenhour1995} but their arguments concerning the size and composition dependence of the heating still apply.

To summarize, we propose that chondritic meteorites form by HIP during the same heating events that form their chondrules. The observed phenomenon of complementarity demonstrates that this at least sometimes happens in a closed system. Atoms released to the gas phase during chondrule formation are trapped by a closed pore structure and/or pressure gradient and recondense as part of the matrix, preserving the chemical and isotopic composition of the original material while redistributing some of it from chondrule to matrix. Compound chondrules and cluster chondrites form in regions of high pre-chondrule density. The temperature, pressure and heating time experienced by the SPP is arguably sufficient for lithification by HIP based on laboratory experience. Detailed modeling and experimentation is clearly needed to verify or refute the hypothesis, but we find it plausible and attractive as an explanation for several properties of chondrules and chondrites that have puzzled cosmochemists for decades.

\subsection{Chondrite Classes, Metamorphic Type and Aqueous Alteration}

It is widely acknowledged that the various classes of meteorites, Enstatite (EC), Ordinary (OC) of H, L or LL subclass, and Carbonaceous (CC) with a number of subclasses, owe their distinctive chemical abundance patterns to their source location within the asteroid belt. Reflectance spectra are used to associate particular meteorite classes with particular asteroids \citep{Burbine2017}. EC's are believed to originate in the inner belt, OC's in the main belt and CC's in the outer belt. Nothing in the flyby model is inconsistent with this interpretation of the data. Distinctive spectral signatures used for classification arise primarily from chemical composition and other factors that are not sensitive to whether the material is in the form of a well lithified chondrite or poorly lithified material. Our model departs from the canonical view of chondrite formation in one main respect: we propose that chondrites were lithified prior to their arrival on a host asteroid, not after. In general, we would favor the view espoused by \citet{Elkins-tanton2011} that host asteroids of chondrites are typically LDPs with chondritic crusts. Most of the chondritic crust, however, would not be in the form of chonrule-laden chondrites but in a lower density, more fragile, primitive and arguably chondrule-free state. 

It is well known that many chondrites experienced metamorphism at temperatures up to 1100 K for thousands to millions of years following their lithification. Following \citet{Elkins-tanton2011}, we suppose that metamorphism occurs on the host LDP within the first few tens of millions of years after the arrival of the chondrite on the asteroid's surface. The heat source for metamorphism is also the radioactive decay of $^{26}$Al, which puts significant constraints on when metamorphism could have occurred \citep{Blackburn2017}. Stochastic variations in surface and shallow depth temperatures result in varying degrees of metamorphosim, generating types 3-7 in the meteorite classification scheme. Once again, the only distinctive feature of the flyby model in this regard is that the lithification of the chondritic meteorites occurred prior to their arrival on the LDP's surface, not after, as is commonly assumed. 

A second observed form of alteration, exhibited by type 1 or 2 meteorites, is aqueous alteration. These objects have hydrated minerals and obviously interacted with water. The canonical view is that this happened on a parent body asteroid. In the flyby model aqueous alteration could occur after the chondrite arrives on the LDP or it could occur within the SPP during the chondrite-forming heating event. \citet{Macke2011} found that porosity declines from type 1 (CI's) through type 4 meteorites. CI's are also the most volatile-rich meteorites and have no chondrules whatsoever. In the context of the flyby model, there is a natural explanation for these properties of CI's - they were simply more distant flybys than other CC's, resulting in gentler heating. In this picture, CI's would represent a transition stage between the well-lithified, chondrule-rich material that appears in our meteorite collections and the lower density, more fragile, less well-lithified material (lacking chondrules) that is not part of those collections, but could be the dominant form of chondritic material on small C-type asteroids. \citet{Sears1998} has discussed a similar point in some detail, arguing that chondrules were much less abundant in the early solar system than the meteorite record, taken at face value, might suggest.     

\subsection{Is the Flyby Mechanism Capable of Processing Enough Mass into Chondrules?}

In our model, chondrules are a by-product of chondrite formation, so the fact that chondrules are ubiquitous within chondrites (except for CI's) is not surprising and does not indicate that chondrules had to be widespread in the Solar System at any time. Chondrites are a minor fraction ($\leq 1$\%) of the primitive solar system material impacting the Earth above the atmosphere at present and there is no reason to think that they, or their chondrules, were originally more common. The flyby model predicts that chondrules will not be found outside of chondrites, a point of view that is consistent with current knowledge according to \citet{Connolly2016}.  

We can roughly estimate the fraction of primitive material still available to us in the solar system today that might have been processed into chondrules and chondrites during their t = 1--4 Myr formation epoch in the following way. Suppose that an SPP makes N$_{fb}$ flybys within 0.1 R$_p$ of one or more LDPs and that the duty cycle for exposed hot lava at any location on the surface of the LDP is given by $\delta$. Then the probability of exposure to chondrule and chondrite forming radiation is N$_{fb} \times \delta$. SPPs that accreted too early would simply be incorporated into the LDP as it fully melted. Those that accreted between 1 and 4 Myr had an increasingly good chance of surviving as part of the growing primitive chondritic crusts. Assuming that the main epoch of accretion for primitive material was over by t = 4 Myr, we may infer that N$_{fb} \ge 1$, since all material that accreted to an LDP had to approach the surface closely at least once. Any primitive material that did not accrete to a larger object could not have survived in the asteroid belt for billions of years and is not part of the fossil record of those times available to us now. Likewise, SPPs that accreted directly into lava oceans lost their primitiveness and may be eliminated from further consideration. 

We can estimate $\delta$ as follows. Surface lava extrusions are prodigious sources of infrared radiation and may well dominate the energy loss of first generation planetesimals. The surface flux of an object with T = 2000 K is $F = \sigma T^4 = 9.1 \times 10^5$ W m$^{-2}$. The energy content of dry dust with a canonical $^{26}$Al/$^{27}$Al ratio of $5 \times 10^{-5}$ is about $6.6 \pm 1$ kJ gm$^{-1}$ according to \citet{Sanders2012}. Adopting this value, the amount of energy available to a 100-200 km radius planetesimal formed near t = 0 with a mean density of 3.3 gm cm$^{-3}$ is $0.91-7.3 \times 10^{26}$ J. If an LDP surface were completely molten at one time, its luminosity would be $1.1--4.6 \times 10^{17}$ W, so it would radiate away its full energy content in a very short period of time, namely 25 -- 50 years. The actual duration of a molten lava extrusion event, however, is likely to be only days and its extent will likely be considerably less than the full surface area of the planetesimal. One would imagine, therefore, a surface similar to what is seen on Io, with episodic eruptions at any given location on some duty cycle, $\delta$. If this state of affairs lasts for 3 Myr, the measured duration of the epoch of chondrule formation, then we can estimate $\delta$ as (30 years) / (3 million years) or $\sim$10$^{-5}$. Of course, LDPs may not radiate all of their available $^{26}$Al energy in this manner so we might expect $\delta$ to be in the range of 10$^{-5}$ - 10$^{-7}$. In this case the fraction of primitive material that had been processed into a well-lithified chondrite and accreted to the chondritic crust of an LDP would be $\gtrsim 10^{-7}$. The rest of the chondritic crust would be material identical in composition to chondrites but not thermally processed into a well-lithified form. 
 
Assuming that most SPPs accrete directly to LDPs without orbiting more than once, we would expect chondrites to represent only a very minor fraction of the primitive crusts of LDPs originally. In the billions of years that follow, however, the fact that they are so well-lithified compared to other primitive material may act to increase their relative abundance in some samples. The rate of arrival on Earth of meteorites in the 10 g to 1 kg mass range, about 85-90\% of which are chondritic \citep{Corrigan2017}, is 2900 - 7300 kg yr$^{-1}$ according to \citet{Bland1996}. \citet{Drolshagen2017} have recently determined the flux of mass reaching the Earth (above the atmosphere) over 34 orders of magnitude in mass, from 10$^{-21}$ to 10$^{12}$ kg, based on a variety of techniques, including in-situ impact data on retrieved space hardware and optical meteor data. They conclude that the best estimate for the mass flux is 54 t per day for an upper size limit of 1 km. The range of values consistent with their data is 30 to 180 t per day, corresponding to 1-6 x 10$^7$ kg yr$^{-1}$, and depends sensitively on the upper limit to the mass size chosen.  Therefore, the well-lithified fraction of primitive material -- chondritic meteorites -- makes up $\sim$10$^{-4}$ of the total mass influx at the top of the Earth's atmosphere. Most of the mass impacting the Earth above the atmosphere is much smaller than the minimum size of chondrites or chondrules, coming in the form of interplanetary dust particles (IDP's). 

Some fraction of the mass impacting the Earth above its atmosphere in all mass ranges must have a cometary origin, of course, but it is uncertain how large that fraction is or how it depends on mass. \citet{Dermott2002} discuss the difficult problem of assigning an origin to the small grains incident at the top of the Earth's atmosphere. They argue that the dominant fraction of IDP's is of asteroidal origin. However, while they have a generally chondritic composition, the IDP's do not resemble ground up fragments of known meteorites. They appear to represent a class of objects quite different from that in our meteorite collections and possibly the major fraction of primitive material that was never thermally processed into chondrules and chondrites. The composition of the asteroid belt is largely unexplored and may consist primarily of this more friable material that does not survive passage to the Earth's surface. While the general agreement between, say, C-type asteroidal spectra and CC meteorite spectra suggests a compositional link, it does not imply that asteroids must contain a large volume fraction of chondrules or that they have been lithified to the extent that chondrites in our meteorite collections have been. It is well known that the mean density of small asteroids that could be parent bodies of chondrites are factors of 2-3 less than chondritic meteorites. This is often interpreted to mean that asteroids are typically ``rubble piles" -- loosley held together fragments of a once denser object -- but they may also simply be made of more porous, friable material than the chondrites. \citet{Sears2017} has recently highlighted evidence that the asteroid Itokawa is not a rubble pile.

To summarize, the flyby model would predict that only a small fraction, perhaps 10$^{-5}$ to 10$^{-7}$ of the material that accretes to larger asteroids as a primitive crust would have been thermally processed into chondrules and chondrites. This may be compared with the measured fraction of primitive material impacting above the Earth's atmosphere that is in the form of chondrites, which we estimate to be in the range of 10$^{-2}$ to 10$^{-4}$. One possible explanation for this mismatch is that SPPs typically make many flybys of LDPs, perhaps commonly orbiting them prior to accretion. \citet{Krot1997} and  \citet{Rubin2017} have discussed cosmochemical evidence of multiple heating events in chondrules. Alternatively, or in addition, there may be other important fractionation processes operating to enhance the proportion of well-lithified to not-well-lithified primitive asteroidal material impacting the Earth. Primitive material must survive for $\sim$4.5 Gyr in the asteroid belt, suffering an enormous number of collisions \citep{Beitz2016}, which may preferentially deplete the fraction that has lesser tensile strength. Small grains ($\le 100 \mu$m) are susceptible to forces from radiation pressure and solar wind that alter their motions and distributions within the Solar System in ways that are different from chondrite-sized objects \citep{Dermott2002}. Cosmic ray exposure ages indicate that OC's typically spend about 10-100 Myr as isolated objects in interplanetary space before impacting the Earth. Tensile strength may be an important factor affecting their probability of survival for that time in that environment. 

\subsection{Relic magnetic fields in chondrules and chondrites}

There is evidence that some chondrules formed in the presence of weak magnetic fields and that chondrites formed or perhaps metamorphosed in weak magnetic fields \citep{Shah2017}. This would be expected in the flyby model because chondrule formation and chondrite lithification both take place very close to the surface of an active LDP. This is followed by accretion to either the same LDP or a different one. As \citet{Elkins-tanton2011} have discussed, objects such as these are expected to have weak magnetic fields due to their liquid cores and strong convection. The existence of evidence for relic magnetism during chondrule formation and/or chondrite metamorphism, therefore, would seem to be entirely consistent with expectation based on the flyby model. 

\subsection{The structure of rocky meteoroids and asteroids}

If, as proposed, the lithification mechanism for chondrites is HIP during a flyby heating, then there will be a maximum mass that can be processed into a monolithic rock, because there is a limited amount of time and energy available. We estimated above that the maximum mass of a chondrite formed in this way is roughly M = $4.7 \times \rho_m^{-2}$ kg, where $\rho_m$ is the density of the material before it is heated. If $\rho_m \ge 10^{-4}$ gm cm$^{-3}$ then the masses of monolithic chondrites should be $M \le 3 \times 10^6$ kg. The largest chondrite recorded in the Meteoritical Bulletin Data Base (URL: https://www.lpi.usra.edu/meteor/metbull.php) is a piece of the Jilin fall, which has a mass of $1.77 \times 10^3$ kg, substantially under this limit \citep{Graham1981}. Most chondrites are smaller than 5 kg and the peak of the mass distribution for those found in Antarctica, where small chondrites are more easily recognized is around 10 gm \citep{Huss1990}.  

A corrollary of this prediction is that rocky meteoroids and asteroids exceeding $\sim10^6$ kg are not monolithic, but composites of chondrites and more porous material. This is consistent with observations of bolides and, in particular, the Chelyabinsk meteoroid, which indicate a mechanical strength for the object as a whole that is two orders of magnitude less than that for the chondrite remnants \citep{Tabetah2017}. Monolithic chondrites exceeding a few thousand tonnes cannot be made by the flyby mechanism so we would further predict that such objects will not be found on any asteroid. There may well be larger boulders that have been lithified by impacts or other mechanisms, but should they exist, we predict they will not be composed primarily of chondrules. We predict that chondrules will be a very minor component of asteroids and, where present, will be confined to well-lithified chondrites similar to those in our museums. This is consistent with the observation that the mean density of small asteroids is significantly less than the mean density of chondrites \citep{Carry2012}. 

\section{Summary}

We propose that chondrules form and chondrites lithify simultaneously when m- to km-scale primitive planetesimals are exposed to infrared radiation from incandescent lava at the surfaces of much larger (100-200 km scale) differentiated planetesimals during close flybys. A thermal model is developed that predicts heating and cooling rates consistent with the constraints of experimental petrology, based on synthetic chondrule textures. The temperature of the extruding lava needs to be $\sim$2000 K, which is reasonable, and the flyby needs to be within $\sim$10\% of the radius of the large planetesimal. The distinctive timescale of $\sim$1 hour for such heating events is a natural consequence of the dynamical timescale of a small object orbiting closely past a larger object with a mean density of around 3 gm cm$^{-3}$. The observed epoch of chondrule formation, t = 1--4 Myr, corresponds to the only time in solar system history when surface lava extrusions should be common: it takes $\sim$1 Myr for sufficiently large planetesimals to form and fully melt, but by $\sim$4 Myr, the power of the $^{26}$Al has subsided sufficiently that crusts should be too thick for frequent ruptures.

In our model, all of the constituents of a chondritic meteorite -- CAI's, pre-solar grains, pre-chondrule and pre-matrix material -- have been assembled during the first 1--4 Myr as an SPP, which survived earlier incorporation into a LDP. The SPP would be quite porous and fragile and chemically and isotopically representative of the location within the solar nebula where it formed. As the object heats during approach to the lava field it will release volatiles, some of which remain trapped within a closed pore structure, to create the relatively high partial pressures of O, Na and Si required of chondrule formation zones. The flyby model predicts a more gradual rise in temperature than the ``flash heating" often assumed for nebular models. Expected temperatures and pressures within the heated planetesimal, 1500 K and 0.1 MPa, are sufficient to plausibly result in lithification by sintering through the HIP process on a timescale of minutes. In some cases, evaporates will remain trapped in pores and recondense as matrix, accounting for the phenomenon of complementarity. In other cases the larger scale melts -- the chondrules -- will squeeze out the pores, resulting in cluster chondrites.

After formation, we propose that these well-lithified chondrites will accrete to larger bodies, where some will experience metamorphosis. In many, if not all, cases the larger body may be the same LDP involved in lithifying the chondrite, so that the ``flyby" event is actually part of the accretion process. Multiple flybys generating multiple heating events prior to final accretion are possible. Once accretion occurs, subsequent evolution is similar to what has been proposed by \citet{Elkins-tanton2011}; metamorphosis occurs rather close to the surface, in zones which are cool enough to be solid, but still warm enough to provide the necessary heat over a longer time scale to account for the observed metamorphic types. Aqueous alteration, responsible for Type 1 and 2 meteorites, on the other hand, plausibly occurs during the pre-accretion heating phase, affecting the more volatile and/or less strongly heated SPPs. Since lithification of the chondrites occurs prior to arrival on the LDP, in our view these objects are more properly called ``host bodies" than  ``parent bodies".

In our scenario, chondrule formation and lithification of the chondrites occur simultaneously, thereby accounting for the fact that chondrules are ubiquitous in almost all classes of chondrites. The exception that proves the rule in this case is the CI's. They lack chondrules entirely, but they are also the least well lithified chondrites. It is likely that only a small fraction of the chondritic crusts of host bodies would be in well-lithified form, because the surface lava eruptions are episodic with a duty cycle of only $\lesssim10^{-5}$ for any particular spot on the surface, so the chance of heating during any particular flyby is small. On the other hand, substantial lithification enhances the likelihood that primitive material will survive to the Earth's surface after a long journey from the asteroid belt. Our meteorite collections contain, therefore, only the rather-well-lithified, and, therefore, chondrule-rich component of the surviving primitive material in the solar system. We predict that chondrules will not be abundant on host asteroids except within chondrites and that chondrites similar to those in our meteorite collections will be a relatively minor component of those asteroids, the rest being more primitive, porous material of lower density and lower tensile strength. 

Our model predicts that there is a size limit to monolithic chondrites that is set by the relatively short timescale of the heating event. Chondrites should in all cases be smaller than $\sim$400 t and, in general, much smaller. We would predict that larger rocky meteorids incident on the Earth would not be monolithic, but composites of individual chondrites embedded in compacted primitive material of lesser tensile strength. This is consistent with the observations that monolithic chondrites larger than 1700 kg have not been found, that most chondrites are quite small (2.5 kg or less) and that rocky meteoroids typically explode on passage through the Earth's atmosphere, sometimes even generating strewn fields composed of meteorites of different classes.  

Laboratory experiments are underway to test whether chondrules with the full range of observed characteristics can be made within the constraints of thermal histories consistent with the flyby model. One prediction of the model is that there should be no measurable age spread among completely melted chondrules in a given meteorite. This is not inconsistent with current data \citep{Alexander2015} but the matter requires further investigation. If relict material can survive within partially melted objects, that could plausibly yield a range in chondrule ages in the flyby model. We further predict that asteroid missions will not reveal host bodies to be monolithic objects -- scaled-up versions of chondrites in our meteorite collections -- but heterogeneous assemblages of much more friable and primitive material with only a small admixture of hardened, high-density rocks. Early press reports from the missions to Ryugu and Bennu appear to support this view although, again, it is too early to know for sure.

If the general explanation for chondrites proposed here is correct, it has implications for our picture of the early solar system. It supports the view that 100-200 km scale planetesimals formed very quickly within the asteroid belt and differentiated due to internal heating by $^{26}$Al. It also suggests that accretion of remaining primitive material occurred over several Myr, creating chondritic crusts on host bodies where the chondrites were initially stored. Collisional fragmentation of host bodies then generated smaller objects that could be transported to near Earth orbits in a variety of ways, generating today's population of meteoroids impacting the Earth. The remaining fraction of those meteoroids, hardened enough to have arrived at Earth and then survive passage through the atmosphere represents our best window on the rather fortunate series of events that allowed them to survive as fossils.      

\acknowledgments

We thank the editor, Alessandro Morbidelli, and referees Guy Libourel, Dominik Hezel and an anonymous referee for helpful comments on an earlier version of this work. We also thank Jim Zareski and Kenichi Abe of Wesleyan University for assistance with the laboratory work. Funding for this project has been provided by NASA, originally through a seed grant from the CT Space Grant Consortium, and with continuing support under award NNX17AE26G. We thank the NSF for support of the SEM at Wesleyan through MRI grant 1725491. Several students have contributed to this work over the years, including Michael Henderson and Tristan Stetson of Wesleyan University and Ian Bania and Katie Chapman of Colgate University. Their participation in the program was funded, in part, by an NSF/REU grant to Wesleyan University supporting the Keck Northeast Astronomy Consortium (AST-1559865) and by the Wesleyan University College of Integrative Studies Summer Research Program. Finally, we are indebted to our colleagues in Wesleyan's Planetary Sciences program, led by Martha Gilmore, for providing the stimulating interdisciplinary environment that has been essential to this work.

\bibliographystyle{apj}
%\bibliography{Icarus.bib}
\bibliography{Collection.bib}

\end{document}